\documentclass[12pt, draftclsnofoot, onecolumn]{IEEEtran}

\usepackage{latexsym,epsf,times,amsmath,color,amssymb,indentfirst,subfigure,fancyhdr,colortbl,bm,caption}
\usepackage{xspace}
\usepackage{graphicx}\epsfxsize=3.in
\usepackage{bbm}
\usepackage{cite}
\usepackage{algorithm}
\usepackage{algorithmic}
\usepackage{amssymb}
\usepackage{amsmath}
\usepackage{cite}
\usepackage{epsfig,graphics,color,amssymb}
\usepackage{epstopdf}
\usepackage{enumerate}

\usepackage{stfloats}
\usepackage{xcolor}
\newtheorem{myTheo}{Theorem}

\newtheorem{myRem}{Remark}

\ifCLASSINFOpdf

\else

\fi

\begin{document}
\title{Weighted Spectral Efficiency Optimization for Hybrid Beamforming in Multiuser Massive MIMO-OFDM Systems}

\author{Jingbo Du,~\IEEEmembership{Student Member,~IEEE,}
		Wei Xu, ~\IEEEmembership{Senior Member, IEEE,}
		\\Chunming Zhao, \emph{Member, IEEE,}
		and Luc Vandendorpe, \emph{Fellow, IEEE}

\thanks{
J. Du is with the National Mobile Communications Research Laboratory, Southeast University, Nanjing 210096, China, and also with the Institute of Information and Communication Technologies, Electronics and Applied Mathematics, Universit\'{e} catholique de Louvain, 1348 Louvain-la-Neuve, Belgium (email: 230159371@seu.edu.cn). W. Xu is with the National Mobile Communications Research Laboratory, Southeast University, Nanjing 210096, China, and also with the Purple Mountain Laboratories, Nanjing 211111, China (email: wxu@seu.edu.cn). C. Zhao is with the National Mobile Communications Research Laboratory, Southeast University, Nanjing 210096, China (email: cmzhao@seu.edu.cn). L. Vandendorpe is with the Institute of Information and Communication Technologies, Electronics and Applied Mathematics, Universit\'{e} catholique de Louvain, 1348 Louvain-la-Neuve, Belgium (e-mail: luc.vandendorpe@uclouvain.be).

Part of this material was presented at the ISWCS 2018. \emph{(Corresponding author: Wei Xu.)}
}% <-this % stops a space
%\thanks{Manuscript received XXX, XX, 2015; revised XXX, XX, 2015.}
}
%\markboth{IEEE Signal Processing Letters}%
%{Submitted paper}

\maketitle

\newtheorem{theorem}{Theorem}
\newtheorem{lemma}{Lemma}
\newtheorem{proposition}{Proposition}

\begin{abstract}
%\boldmath
In this paper, we consider hybrid beamforming designs for multiuser massive multiple-input multiple-output (MIMO)-orthogonal frequency division multiplexing (OFDM) systems. Aiming at maximizing the weighted spectral efficiency, we propose one alternating maximization framework where the analog precoding is optimized by Riemannian manifold optimization. If the digital precoding is optimized by a locally optimal algorithm, we obtain a locally optimal alternating maximization algorithm. In contrast, if we use a weighted minimum mean square error (MMSE)-based iterative algorithm for digital precoding, we obtain a suboptimal alternating maximization algorithm with reduced complexity in each iteration. By characterizing the upper bound of the weighted arithmetic and geometric means of mean square errors (MSEs), it is shown that the two alternating maximization algorithms have similar performance when the user specific weights do not have big differences. Verified by numerical results, the performance gap between the two alternating maximization algorithms becomes large when the ratio of the maximal and minimal weights among users is very large. Moreover, we also propose a low-complexity closed-form  method without iterations. It employs matrix decomposition for the analog beamforming and weighted MMSE for the digital beamforming. Although it is not supposed to maximize the weighted spectral efficiency, it exhibits small performance deterioration compared to the two iterative alternating maximization algorithms and it qualifies as a good initialization for iterative algorithms, saving thereby iterations.
\end{abstract}


\IEEEpeerreviewmaketitle
\section{Introduction}
Multiple-input multiple-output (MIMO) is a common technology improving the capacity of a radio link using multiple transmit and receive antennas. In future wireless systems, the base station (BS) can be equipped with hundreds, or even thousands, of antennas, i.e., namely massive MIMO. As an extension of MIMO, massive MIMO exhibits its capability of achieving a much larger array gain and more aggressive spatial multiplexing than conventional MIMO \cite{5595728, 6375940, 6736761}. However, in conventional MIMO systems, fully-digital beamforming requires that each antenna be driven by a dedicated radio frequency (RF) chain which is not suitable for massive MIMO anymore. The growing number of RF chains remarkably increases hardware cost, system complexity and power consumption which causes challenges in system implementation. To overcome this problem, hybrid analog and digital beamforming architectures have been proposed, where RF chains are reused \cite{6717211, 7397861, 7160780, 6928432, 7947159} by making several antennas share fewer RF chains. In short, hybrid beamforming consists of a high dimensional analog beamformer and a low-dimensional digital beamformer. Most popularly, a low-cost phase shifter network is applied to implement the analog beamforming to achieve a tradeoff between degrees of freedom (DOF) and system implementational complexity.

Hybrid beamforming designs have been considered for single-user MIMO systems \cite{6717211, 7397861} and multiuser MIMO systems \cite{7160780, 6928432, 7947159} over single-carrier channels. However, hybrid beamforming designs over frequency selective channels are more challenging. The analog beamforming is implemented by analog components in the time domain, which implies that it can only be ``frequency flat''. In other words, the designs for narrowband hybrid beamforming are not applicable anymore in orthogonal frequency division multiplexing (OFDM) systems because the analog beamforming is the same across all subcarriers. Moreover, there are numerous studies on multiuser digital precoding but much fewer contributions on multiuser analog precoding which makes analog precoding part in hybrid precoding more attractive.

Generally, there are three popular types of methods for the OFDM-based analog precoding implemented by the phase shifter network. One is relaxing the constant amplitude constraints \cite{7880698, 7737056, 8352108}; a second one is based on specific codebooks \cite{7448873}; a last one directly solves the problem using iterative methods free from the restriction of the codebook \cite{7397861, 7913599}. The first two types of methods reduce the computational complexity at the cost of some performance deterioration. The alternating minimization algorithm proposed in \cite{7397861} is designed for the single-user case and not suitable for the multiuser case. If the alternating minimization algorithm is used in multiuser massive MIMO, the algorithm first calculates the optimal fully-digital beamforming solution and then minimizes the difference between the optimal fully-digital beamforming and the hybrid beamforming. However, it is difficult to calculate the optimal multiuser solution especially for massive MIMO which requires quite large complexity. On the other hand, frequency flat precoding and combining are sufficient to achieve the maximum spectral efficiency in single-user systems when there is not too much scattering in the channel as shown in \cite{7887656}. Therefore, approaching the optimal fully-digital precoding by hybrid precoding with frequency flat analog precoding is effective in single-user systems over millimeter wave channels. However, interference among users, which does not exist in the single-user case, plays an important role in the system performance in multiuser systems. Generally, the accuracy of approaching fully-digital precoding by hybrid precoding falls with the number of subcarriers and results in more multiuser interference. Performance reduction follows multiuser interference which makes alternating minimization not suitable for OFDM-based multiuser hybrid beamforming anymore. The proposed heuristic algorithm in \cite{7913599} is based on the average of the covariance matrices of frequency domain channels which reduces computational complexity but causes performance deterioration.

In this paper, we plan to propose algorithms to maximize the weighted spectral efficiency for OFDM-based hybrid beamforming in multiuser massive MIMO. They do not need the relaxation of the constant amplitude constraints and are free from the restriction of codebooks. To tackle this challenge, we propose one alternating maximization framework with two iterative algorithms. In each iteration, the analog precoder and digital precoder are alternately optimized to maximize the weighted spectral efficiency while keeping the other one fixed. Iterations are repeated until convergence. Moreover, we also propose a low-complexity closed-form multiuser hybrid beamforming design for OFDM systems without iterations. For real-world systems, we investigate the performance of three proposed designs based on imperfect channel information with random error. It is revealed that the performance does not reduce a lot compared to the case with perfect channel information. Note that this paper is an extension of \cite{8491217} which only considers unweighted spectral efficiency. Compared to algorithms in \cite{8491217}, we propose new ones, including a locally optimal algorithm, for a more general system and we provide more discussions and insights on the convergence, computational complexities and system performance. In particular, we remove the assumption that the number of RF chains is equal to that of users and consider weighted spectral efficiency as the objective function. More specifically, the main contributions of this paper are summarized as follows:

%\begin{itemize}

%\item
(1) We introduce an alternating maximization framework for OFDM-based multiuser hybrid precoding. It uses Riemannian manifold optimization for the analog precoding and different digital beamforming designs. Different from the Riemannian manifold minimization in \cite{7397861}, the proposed Riemannian manifold maximization algorithm directly solves the weighted spectral efficiency maximization problem to achieve a local optimum for the analog precoding.

%\item
(2) For digital beamforming, we utilize a locally optimal design and a weighted MMSE-based one. Combining the Riemannian manifold optimization and the locally optimal digital beamforming, we get a locally optimal alternating maximization algorithm for OFDM-based multiuser hybrid precoding. Replacing the locally optimal digital precoding by weighted MMSE-based digital precoding, we get another alternating maximization algorithm whose performance is shown to be similar with the locally optimal hybrid beamforming, especially when the weights among users are similar.

%\item
(3) A low-complexity closed-form design without iterations is also proposed for the OFDM-based hybrid beamforming which uses a channel matrix decomposition design for the analog precoding and weighted MMSE for the digital precoding. Although this approach is not supposed to maximize system performance, it is verified by simulation results that its performance only suffers from a small performance degradation compared to that of the alternating optimization algorithms. In addition, this approach is also a good initialization of the iterative algorithm, thereby saving iterations.

%\end{itemize}

The rest of the paper is organized as follows. System and channel models are first introduced in Section II. We then propose two alternating maximization algorithms to maximize the weighted spectral efficiency in Section III and a simplified design with reduced complexity in Section IV. Simulation results are presented in Section V before concluding remarks in Section VI.

Notations throughout this paper are as follows. Upper and lower case bold-face letters are matrices and vectors, respectively. Italicized variables are scalars. $\|\cdot\|_F$ and $(\cdot)^H$ represent the Frobenius norm and Hermitian of a matrix, respectively. $|\cdot|$ and $(\cdot)^*$ represent the absolute value and conjugation of a complex number, respectively. $\simeq$ is used to indicate asymptotically equal to. $\mathbb{C}^{m\times n}$ and $\mathbb{R}^{m\times n}$ respectively denote the ensemble of complex and real valued $m\times n$ matrices. $\mathbf{I}$ stands for the identity matrix. $\mathrm{Tr}[\cdot]$ refers to the trace. $\mathbb{E}[\cdot]$ is used to denote expectation. $\Re[\cdot]$ represents the real part of a complex number. $\mathcal{CN}(\mu,\sigma^2)$ stands for the complex Gaussian distribution with mean $\mu$ and variance $\sigma^2$. $\mathcal{U}(a,b)$ denotes the uniform distribution between $a$ and $b$. $\circ$ refers to the Hadamard product. ${\rm diag}(a_1,...,a_m)$ refers to a diagonal matrix whose diagonals are $a_1,...,a_m$.

\section{System and Channel Models}
\subsection{System Model}
We consider an OFDM-based massive MIMO hybrid beamforming wireless communication system. The base station (BS) is equipped with $N_{RF}$ RF chains and $M$ antennas where $N_{RF}\leq M$. % as shown in Fig. 1.
For OFDM, $K$ subcarriers are used for data transmission where $U$ single-antenna users are simultaneously served on the entire band. At the BS, the signals at different subcarriers are first digitally-precoded respectively and then transformed to the time domain by using $K$-point inverse fast Fourier transforms (IFFTs). After that, the transformed signals are finally processed by an analog precoding matrix before transmission through the antenna array. Since the analog precoder is operated on signals after IFFT, it is the same for all subcarriers, which indicates that it is flat in the frequency domain. This is the key challenge in designing the hybrid precoding over frequency selective wideband channels.

%\begin{figure}[!htpb]
%\centering
%\begin{minipage}{1\textwidth}
%\centering
%\includegraphics[width=0.9\textwidth]{fig1.eps}
%\caption{OFDM-based Hybrid precoding architecture.}
%\end{minipage}
%\end{figure}

%The signal transmitted at subcarrier $k$ can be represented as
%\begin{align}
%\mathbf{x}[k]=\mathbf{F}\mathbf{W}[k]\mathbf{s}[k]\label{processedsignal}
%\end{align}
Assuming a block-fading channel model, the signal received at subcarrier $k$ can be written by
\begin{align}
\mathbf{y}[k]=&\mathbf{H}[k]^H\mathbf{F}\mathbf{W}[k]\mathbf{s}[k]+\mathbf{n}[k]\label{receivedsignal}
\end{align}
where $\mathbf{s}[k]\in\mathbb{C}^{U\times 1}$ denotes the vector of transmitted data symbols at subcarrier $k$ with $\mathbb{E}\{\mathbf{s}[k]\mathbf{s}^H[k]\}=\mathbf{I}$, $\mathbf{W}[k]=[\mathbf{w}_1[k],\mathbf{w}_2[k],...,\mathbf{w}_U[k]]\in\mathbb{C}^{N_{RF}\times U}$ refers to the digital precoder at subcarrier $k$, $\mathbf{F}=[\mathbf{f}_1,\mathbf{f}_2,...,\mathbf{f}_{N_{RF}}]\in\mathbb{C}^{M\times N_{RF}}$ stands for the analog precoder, $\mathbf{H}[k]=[\mathbf{h}_1[k],\mathbf{h}_2[k],...,\mathbf{h}_U[k]]\in\mathbb{C}^{M\times U}$ represents the channel matrix and $\mathbf{n}[k]\sim\mathcal{CN}(\mathbf{0},\sigma_n^2\mathbf{I})$ refers to the additive white Gaussian noise at subcarrier $k$ in which $\sigma_n^2$ is the noise power. It is notable that the analog part of the hybrid beamformer is typically implemented using simple analog components such as analog phase shifters which can only modify the angles of signals. Thus, every entry in $\mathbf{F}$ has the same constant amplitude. In this work, the fully-connected structure for hybrid precoding is considered in which each RF chain drives all antennas. Each RF chain and each antenna is connected through only one phase shifter. Therefore, the $i$-th element of $\mathbf{f}_n$ is normalized as
\begin{align}
|f_{ni}|=\frac{1}{\sqrt{M}}.
\end{align}

%Assuming a block-fading channel model, the signal received at subcarrier $k$ can be written by
%\begin{align}
%\mathbf{y}[k]=&\mathbf{H}[k]^H\mathbf{x}[k]+\mathbf{n}[k]\label{receivedsignal}
%\end{align}
%where $\mathbf{H}[k]=[\mathbf{h}_1[k],\mathbf{h}_2[k],...,\mathbf{h}_U[k]]\in\mathbb{C}^{M\times U}$ represents the channel matrix and $\mathbf{n}[k]\sim\mathcal{CN}(\mathbf{0}_K,P_n\mathbf{I}_K)$ refers to the additive white Gaussian noise for subcarrier $k$ in which $P_n$ is the noise power.

From \eqref{receivedsignal}, the received signal of the $u$-th user at subcarrier $k$ is
\begin{align}
y_u[k]=&\mathbf{h}_u[k]^H\mathbf{F}\mathbf{W}[k]\mathbf{s}[k]+n_u[k]\nonumber\\
=&\mathbf{h}_u[k]^H\mathbf{F}\mathbf{w}_u[k]\mathbf{s}_u[k]+\mathbf{h}_u[k]^H\mathbf{F}\sum\limits_{i\neq u}\mathbf{w}_i[k]\mathbf{s}_i[k]+n_u[k]
\end{align}
where $y_u[k]$, $s_u[k]$ and $n_u[k]$ respectively denote the $u$-th element of $\mathbf{y}[k]$, $\mathbf{s}[k]$ and $\mathbf{n}[k]$.

\subsection{Channel Model}
To characterize the scattering features of mmWave channels, we adopt the most widely used geometric channel model \cite{6387266}. We assume that the channel between the BS and the $u$-th user has $C_u$ clusters and each cluster has $L_{uc}$ scatterers. For the $u$-th user, each cluster has a time delay $\tau_{uc}\in\mathbb{R}$. For the $c$-th cluster of the $u$-th user, each scatterer has an angle of departure (AOD) $\theta_{ucl}\in[0,2\pi]$ and complex path gain $\alpha_{ucl}$. Assuming sampling with period $T$, the $d$-th delay tap for the link between BS and the $u$-th user follows the expression
\begin{align}
\mathbf{h}_{ud}=\sqrt{\frac{M}{C_uL_u}}\sum_{c=1}^{C_u}\left[p(dT-\tau_{uc})\sum_{l=1}^{L_{uc}}\alpha_{ucl}\mathbf{a}(\theta_{ucl})\right]\label{chtd}
\end{align}
where $p(\tau)$ is the pulse shaping filter and $\mathbf{a}(\theta_{ucl})$ denotes the antenna array response vector of the base station. In this work, uniform linear arrays (ULAs) are utilized where
\begin{align}
\mathbf{a}(\theta_{ucl})=\frac{1}{\sqrt{M}}[1,\mathrm{e}^{-j\frac{2\pi}{\lambda}\delta\sin(\theta_{ucl})},...,\mathrm{e}^{-j(M-1)\frac{2\pi}{\lambda}\delta\sin(\theta_{ucl})}]^T
\end{align}
in which $\lambda$ denotes the signal wavelength, and $\delta$ represents the distance between adjacent antenna elements. In the frequency domain, we write the channel vector for the $u$-th user as
\begin{align}
\mathbf{h}_u[k]=\sum_{d=0}^{D-1}\mathbf{h}_{ud}\mathrm{e}^{-j\frac{2\pi kd}{K}}\label{eqcmintd}
\end{align}
where $D$ refers to the number of delay taps. Note that most of the results developed in this paper are general for massive MIMO channels, and not restricted to the channel model in this subsection. We describe the mmWave channel model here as it will be adopted for simulations in Section V.

\subsection{Problem Formulation}
The problem of interest for the multiuser MIMO case is to design the hybrid analog and digital beamformers in order to maximize the weighted spectral efficiency. In this case, the spectral efficiency of the $u$-th user at subcarrier $k$ can be formulated as
\begin{align}
R_u[k]=&\log_2\left(1+\frac{|\mathbf{h}_u[k]^H\mathbf{F}\mathbf{w}_u[k]|^2}{\sum\limits_{i\neq u}|\mathbf{h}_u[k]^H\mathbf{F}\mathbf{w}_i[k]|^2+\sigma_n^2}\right)\nonumber\\
=&\log_2\left(1+\frac{\vartheta|\mathbf{h}_u[k]^H\mathbf{F}\mathbf{w}_u[k]|^2}{\vartheta\sum\limits_{i\neq u}|\mathbf{h}_u[k]^H\mathbf{F}\mathbf{w}_i[k]|^2+1}\right)\label{se}
\end{align}
where $\vartheta=\frac{1}{\sigma_n^2}$ refers to the transmitted signal power relative to the noise power. For the system, the problem is written as
\begin{align}
\max\limits_{\mathbf{F},\{\mathbf{W}[k]\}_{k=1}^K}\quad &\sum\limits_{k=1}^K\sum\limits_{u=1}^Uz_uR_u[k]\label{pro}\\
\mathrm{s.t.}\qquad &|f_{ni}|=\frac{1}{\sqrt{M}},1\leq i\leq M,\ 1\leq n\leq N_{RF} \label{anacon}\\
 &\|\mathbf{F}\mathbf{w}_u[k]\|_F=1,1\leq u\leq U,\ 1\leq k\leq K\label{digcon}
\end{align}
where $0<z_u<1$ is the spectral efficiency weighting factor of the $u$-th user and $\sum\limits_{u=1}^Uz_u=1$. To tackle this problem, we would like to propose some design strategies in the following. Note that perfect channel state information (CSI) is assumed to be available in order to study the performance limits of the hybrid beamforming architecture. In massive MIMO, although channel estimation could be challenging in practice, there have been some schemes developed in \cite{7961152, 8323164} for OFDM-based hybrid architecture.

In practical applications, the weighting factors may not satisfy $\sum\limits_{u=1}^Uz_u=1$. For example, there may be a system serving $U$ users with unconstrained positive weights $l_1,l_2,...,l_U$. To fit our work to this system, we can normalize the weights as $z_u=\frac{l_u}{\sum\limits_{i=1}^{U}l_i}$. After designing the hybrid beamforming, the weighted spectral efficiency should be rewritten as $R_{\rm psum}=R_{\rm sum}\sum\limits_{i=1}^{U}l_i=\left(\sum\limits_{k=1}^K\sum\limits_{u=1}^Uz_uR_u[k]\right)\left(\sum\limits_{i=1}^{U}l_i\right)$.

\section{Alternating Maximization Algorithm for Hybrid Precoding}

In this section, we would like to propose some iterative algorithms for problem \eqref{pro} based on alternating optimization. As apparent from \eqref{pro}, the main difficulty of this problem is solving the maximization problem over two sets of variables, i.e., $\mathbf{F}$ and $\{\mathbf{W}[k]\}_{k=1}^K$. However, alternating optimization algorithms can be used in such applications thanks to their iterative nature and simplicity. In this section, we introduce one alternating maximization framework for hybrid precoding to maximize the weighted spectral efficiency. Instead of solving the original optimization problem over two sets of variables, the proposed alternating maximization framework respectively maximizes the weighted spectral efficiency with respect to $\mathbf{F}$ and $\{\mathbf{W}[k]\}_{k=1}^K$ while keeping the other one fixed. In this framework, we have one Riemannian manifold optimization algorithm for the analog precoding and two different digital precoding designs which further lead to two alternating maximization algorithm for multiuser hybrid precoding.

\subsection{Riemannian Manifold Optimization for Analog Precoding}

To design the analog precoding with fixed digital precoding, problem \eqref{pro} becomes
\begin{align}
\max\limits_{\mathbf{F}}\quad &\sum\limits_{k=1}^K\sum\limits_{u=1}^Uz_uR_u[k]\label{anapro}\\
\mathrm{s.t.}\quad &\eqref{anacon}.\nonumber
\end{align}
Unfortunately, the non-convex constant amplitude constraints still make the problem difficult to solve. To the best of the authors' knowledge, there is no general approach to solve \eqref{anapro} optimally. In this subsection, we would like to propose an effective Riemannian manifold maximization algorithm which directly solves \eqref{anapro} to find a locally optimal solution. In the following, we first introduce some definitions for manifold optimization before proposing the locally optimal algorithm.

In mathematics, a manifold is a topological space that locally resembles Euclidean space near each point. If we transfer $\mathbf{F}$ into a vector $\mathbf{x}=\mathrm{vec}[\mathbf{F}]$, we consider an embedded submanifold of the Euclidean space $\mathbb{C}^{MN_{RF}}$ for the analog precoding as
\begin{align}
\mathcal{M}=\{\mathbf{x}\in\mathbb{C}^{MN_{RF}}:|x_i|=\frac{1}{\sqrt{M}}, i=1,2,...,MN_{RF}\}.
\end{align}
where $x_i$ is the $i$-th element of $\mathbf{x}$. Normally, the manifold $\mathcal{M}$ is not as friendly as the Euclidean or the vector space for an optimization problem. To address this issue, the Riemannian manifold is defined as a manifold whose tangent spaces are endowed with a smoothly varying inner product. In the same way that the derivative of a complex-valued function provides a local linear approximation of the function, the tangent space $\mathcal{T}_\mathbf{x}\mathcal{M}$ at point $\mathbf{x}$ provides a local vector space approximation of the manifold $\mathcal{M}$. Moreover, the inner product also makes it possible to define various geometric notions on the manifold. Then, by treating $\mathbb{C}$ as $\mathbb{R}^2$ with the canonical inner product, we define the Euclidean metric in the complex $\mathbb{C}$ plane as
\begin{align}
<x_1,x_2>=\Re[x_1x_2^*].
\end{align}
The definition of inner product allows us to define the tangent vector of $\mathcal{M}$. For a vector $\mathbf{z}$, it is considered as orthogonal to $\mathbf{x}$ if the inner product of every element in $\mathbf{z}$ and the corresponding element in $\mathbf{x}$ equals to zero, i.e.,
\begin{align}
<z_i,x_i>=0,\forall i
\end{align}
where $z_i$ is the $i$-th element of $\mathbf{z}$. By treating each complex element in the vectors as a vector in $\mathbb{R}^2$, we can state that $\mathbf{z}$ is a tangent vector of $\mathcal{M}$ at $\mathbf{x}$ if
\begin{align}
\Re[\mathbf{z}\circ\mathbf{x}^*]=\mathbf{0}.
\end{align}
The tangent space to a manifold  $\mathcal{M}$ at point $\mathbf{x}$, denoted by $\mathcal{T}_\mathbf{x}\mathcal{M}$, is the set of all tangent vectors to $\mathcal{M}$ at $\mathbf{x}$. Therefore, the tangent space of $\mathbf{x}$ is denoted as
\begin{align}
\mathcal{T}_\mathbf{x}\mathcal{M}=\{\mathbf{z}\in\mathbb{C}^{MN_{RF}}:\Re[\mathbf{z}\circ\mathbf{x}^*]=\mathbf{0}\}.
\end{align}

Due to the fact that the neighborhood of each point on a manifold resembles the Euclidean space \cite{lee2001introduction}, optimization algorithms in the Euclidean space can also be locally applied over the Riemannian manifold. As the tangent space provides a friendly vector space for the optimization problem, some line search methods can be employed. Thanks to this fact, we would like to propose a conjugate gradient algorithm for analog precoding based on Riemannian manifold optimization in the following.

According to the vectorized analog precoder, we define the cost function as
\begin{align}
f(\mathbf{x})\triangleq R_{\rm sum}.
\end{align}

To maximize the weighted spectral efficiency, the direction of the greatest increase of the cost function on the tangent space of the current point is needed. In the Riemannian manifold optimization, the Euclidean gradient is first obtained by using the fact that $\frac{\partial f(\mathbf{x})}{\partial\mathbf{x}^*}$ is the vectorized Euclidean gradient of $\mathbf{F}$, i.e., $\frac{\partial f(\mathbf{x})}{\partial\mathbf{x}^*}=\mathrm{vec}\left[\frac{\partial f(\mathbf{x})}{\partial\mathbf{F}^*}\right]$ where
\begin{align}
\frac{\partial f(\mathbf{x})}{\partial\mathbf{F}^*}=&\sum\limits_{k=1}^K\sum\limits_{u=1}^U\frac{z_u}{\ln2}\left[\frac{\vartheta\mathbf{h}_u[k]\mathbf{h}_u[k]^H\mathbf{F}\mathbf{W}[k]\mathbf{W}[k]^H}{1+\vartheta\|\mathbf{h}_u[k]^H\mathbf{F}\mathbf{W}[k]\|_F^2}-\frac{\vartheta\mathbf{h}_u[k]\mathbf{h}_u[k]^H\mathbf{F}\mathbf{W}_{\bar{u}}[k]\mathbf{W}_{\bar{u}}[k]^H}{1+\vartheta\|\mathbf{h}_u[k]^H\mathbf{F}\mathbf{W}_{\bar{u}}[k]\|_F^2}\right]\label{anader}
\end{align}
in which $\mathbf{W}_{\bar{u}}[k]$ consists of all columns of $\mathbf{W}[k]$ except for $\mathbf{w}_u[k]$. The proof for \eqref{anader} can be found in Appendix A. Then the Euclidean gradient is forced onto the tangent space with the orthogonal
projection as
\begin{align}
\mathrm{P}_{\mathbf{x}}\left(\frac{\partial f(\mathbf{x})}{\partial\mathbf{x}^*}\right)=\frac{\partial f(\mathbf{x})}{\partial\mathbf{x}^*}-\Re\left[\frac{\partial f(\mathbf{x})}{\partial\mathbf{x}^*}\circ\mathbf{x}^*\right]\circ\mathbf{x}.
\end{align}

If the concept of moving in the direction of a vector is straightforward as in the Euclidean space, the point $\mathbf{x}$ moves in the tangent space $\mathcal{T}_\mathbf{x}\mathcal{M}$. On a manifold, the notion of a retraction mapping generalizes the notion of moving in the direction of a tangent vector where $\mathbf{x}$ stays on the manifold $\mathcal{M}$. Given $\mathbf{x}$ in $\mathcal{M}$, the search step $\alpha$ and the search direction $\mathbf{d}$, compute the new point as
\begin{align}
\mathrm{Ret}_{\mathbf{x}}(\alpha\mathbf{d})=\frac{1}{\sqrt{M}}\mathrm{vec}\left[\frac{x_1+\alpha d_1}{|x_1+\alpha d_1|},...,\frac{x_{MN_{RF}}+\alpha d_{MN_{RF}}}{|x_{MN_{RF}}+\alpha d_{MN_{RF}}|}\right].
\end{align}

With the operations of calculating the gradient and moving the point, we develop the conjugate gradient algorithm for the analog precoding as shown in Algorithm 1.
\begin{algorithm}
\caption{Conjugate Gradient Algorithm for Analog Precoding Based on Manifold Optimization}
\begin{algorithmic}[1]
\REQUIRE $\mathbf{W}[k]_{k=1}^K$, $\mathbf{x}_0$
\STATE $\mathbf{d}_0=\mathrm{P}_{\mathbf{x}_{0}}\left(\frac{\partial f(\mathbf{x})}{\partial \mathbf{x}^*}\bigg|_{\mathbf{x}=\mathbf{x}_{0}}\right)$, $t=0$ and $f(\mathbf{x}_{-1})=0$
\REPEAT
\STATE Choose Armijo backtracking line search step size $\alpha_t$ \cite[Definition 4.2.2]{absil2009optimization}
\STATE Update the position as $\mathbf{x}_{t+1}=\mathrm{Ret}_{\mathbf{x}_t}(\alpha_t\mathbf{d}_t)$
\STATE Compute Riemannian gradient: $\mathbf{g}_{t+1}=\mathrm{P}_{\mathbf{x}_{t+1}}\left(\frac{\partial f(\mathbf{x})}{\partial\mathbf{x}^*}\bigg|_{\mathbf{x}=\mathbf{x}_{t+1}}\right)$
\STATE Calculate Polak-Ribiere parameter as $\beta_{t+1}=\frac{\mathbf{g}_{t+1}^H(\mathbf{g}_{t+1}-\mathrm{P}_{\mathbf{x}_{t+1}}(\mathbf{g}_{t}))}{\|\mathrm{P}_{\mathbf{x}_{t+1}}(\mathbf{g}_{t})\|_F^2}$
\STATE Determine conjugate direction: $\mathbf{d}_{t+1}=\mathbf{g}_{t+1}+\beta_{t+1}\mathrm{P}_{\mathbf{x}_{t+1}}(\mathbf{d}_t)$
\STATE $t\leftarrow t+1$
\UNTIL{$(f(x_{t})-f(x_{t-1}))/f(x_{t-1})<\omega$ for a small $\omega>0$}
\end{algorithmic}
\end{algorithm}

Note that we employ Armijo backtracking to choose the step size which is calculated as $\alpha_t=ab^m$ where $m$ is the smallest integer such that
\begin{align}
f(\mathbf{x}_t)-f\left(\mathrm{Ret}_{\mathbf{x}_t}\left(ab^m\mathbf{d}_t\right)\right)\geq -cab^m||\mathbf{d}_t||_F^2
\end{align}
where $c>0$, $a$ and $b$ are fixed scalars between zero and one. According to Theorem 4.3.1 in \cite{absil2009optimization}, Algorithm 1 is guaranteed to converge to a critical point.

In this section, we propose a gradient-based method to optimize elements of $\mathbf{F}$. It is notable that the phases of elements can also be optimized in terms of spectral efficiency. To optimize the phases $\theta_{ni}$, we first write the gradients with respect to phases as $\frac{\partial f(\mathbf{x})}{\partial\theta_{ni}}=\frac{\partial f(\mathbf{x})}{\partial f_{ni}^*}\frac{\partial f_{ni}^*}{\partial\theta_{ni}}=\frac{-j\exp(j\theta_{ni})}{\sqrt{M}}\frac{\partial f(\mathbf{x})}{\partial f_{ni}^*}$ where $\frac{\partial f(\mathbf{x})}{\partial f_{ni}^*}$ is the $(n,i)$-th element of $\frac{\partial f(\mathbf{x})}{\partial\mathbf{F}^*}$. $\frac{\partial f(\mathbf{x})}{\partial\theta_{ni}}$ is probably complex which implies that we still need the retraction mapping operation to map the updated point, i.e. $(\theta_{ni})_{t+1}$, back onto the manifold. Therefore, we still need to use Riemannian manifold optimization. Moreover, the updated point is possibly out of $[0,2\pi)$. According to $\exp(j\eta)=\exp(j(\eta+2m\pi)), \forall m\in\mathbb{Z}$, it is not difficult to map $(\theta_{ni})_{t+1}$ back to $[0,2\pi)$. However, it becomes difficult to operate orthogonal projection. Although gradients with respect to $\mathbf{\theta}_{ni}+2m\pi,\forall m\in\mathbb{Z}$ are the same, the tangent spaces for orthogonal projection become different. In this case, some conclusions require additional proof which is difficult at the present. For example, the solution to the gradient-based algorithm optimizing phases is not guaranteed to be locally optimal. Therefore, we still suggest optimizing constant magnitude elements instead of phases.

\subsection{Digital Precoding Designs}

If the receiver of the $u$-th user at subcarrier $k$ is designed according to MMSE, i.e.,
\begin{align}
b_u[k]=&\left(\mathbf{g}_u[k]^H\mathbf{W}[k]\mathbf{W}[k]^H\mathbf{g}_u[k]+\frac{1}{\vartheta}\right)^{-1}\mathbf{g}_u[k]^H\mathbf{w}_u[k],\label{receiver}
\end{align}
the signal received by user $u$ at subcarrier $k$ can be rewritten as
\begin{align}
y_u[k]=&b_u[k]^H\left(\mathbf{g}_u[k]^H\mathbf{W}[k]\mathbf{s}[k]+n_u[k]\right)\label{sigre}
\end{align}
where the effective channel at subcarrier $k$ is represented by
\begin{align}
\mathbf{G}[k]^H&=\mathbf{H}[k]^H\mathbf{F}.\label{effectivech}
\end{align}
in which $\mathbf{G}[k]=[\mathbf{g}_1[k],...,\mathbf{g}_U[k]]$. With the receiver, the MSE of the $u$-th user at subcarrier $k$ is given by
\begin{align}
\xi_u[k]=&\mathbb{E}\left[(y_u[k]-s_u[k])(y_u[k]-s_u[k])^*\right]\nonumber\\
=&b_u[k]^*\left(\mathbf{g}_u[k]^H\mathbf{W}[k]\mathbf{W}[k]^H\mathbf{g}_u[k]+\frac{1}{\vartheta}\mathbf{I}\right)b_u[k]\nonumber\\
&-b_u[k]^*\mathbf{g}_u[k]^H\mathbf{w}_u[k]-\mathbf{w}_u[k]^H\mathbf{g}_u[k]b_u[k]+1\nonumber\\
\overset{(a)}{=}&1-\mathbf{w}_u[k]^H\mathbf{g}_u[k]\left(\mathbf{g}_u[k]^H\mathbf{W}[k]\mathbf{W}[k]^H\mathbf{g}_u[k]+\vartheta^{-1}\mathbf{I}\right)^{-1}\mathbf{g}_u[k]^H\mathbf{w}_u[k]
\end{align}
where (a) uses \eqref{receiver}. In this case, we have the well-known relation between the achieved SINR, denoted as $\mathrm{SINR}_u[k]$, and $\xi_u[k]$ \cite{7264975, 4509444, 532892}:
\begin{align}
\mathrm{SINR}_u[k]=\frac{1}{\xi_u[k]}-1.
\end{align}
Then, the spectral efficiency of the $u$-th user at subcarrier $k$ can also be expressed as $R_u[k]=\log_2\frac{1}{\xi_u[k]}$.

%By this way, the spectral efficiency maximization problem for the digital beamforming can be formulated as
%\begin{align}
%\max\limits_{\{\mathbf{W}[k]\}_{k=1}^K}\quad &\sum\limits_{k=1}^K\sum\limits_{u=1}^Uz_u\log_2\frac{1}{\xi_u[k]}\\
%\mathrm{s.t.}\qquad &\eqref{digcon}.\nonumber
%\end{align}

Since $\mathbf{W}[k]$ is only related to the spectral efficiency at subcarrier $k$, the design for digital precoding can be divided into $K$ independent problems. After straightforward mathematical manipulations, the digital beamforming design problem can be equivalently expressed as
\begin{align}
\min\limits_{\mathbf{W}[k]}\quad &\prod\limits_{u=1}^U(\xi_u[k])^{z_u}\label{mmsepro}\\
\mathrm{s.t.}\qquad &\|\mathbf{F}\mathbf{w}_u[k]\|_F\leq1,\forall u\label{digconuser}.
\end{align}

This problem is quite similar to the narrowband digital precoding design except for the constraints \eqref{digconuser}. It implies that some pure digital precoding designs could be applied based on the effective channels. We first use a locally optimal digital precoding as in \cite{6104172}. The algorithm in \cite{6104172} uses an iterative algorithm to design the digital precoding. In each iteration of the locally optimal digital precoding algorithm, we first calculate the receiver as \eqref{receiver} and update some necessary factors according to
\begin{align}
\gamma_u=&\frac{1}{1-z_u},\mu_u=\frac{1}{z_u}-1,\kappa_u=z_u\mu_u^{1-z_u},
\nu_u[k]=\frac{\left[\prod\limits_{i=1}^U\xi_i[k]^{z_u}\right]^{\frac{1}{U}}}{\xi_u[k]^{z_u}},\zeta_u[k]=\left[\frac{\nu_u[k]^{\gamma_u}}{\mu_u\xi_u[k]}\right]^{\frac{1}{\mu_u+1}}\label{factors}
\end{align}
with fixed $\mathbf{W}[k]$. More details on \eqref{factors} are included in Appendix B. In the second step, the digital precoding $\mathbf{W}[k]$ is optimized by solving a second-order cone program (SOCP) problem
\begin{align}
\min\limits_{\chi,\mathbf{W}[k]}\quad&\chi\label{socp}\\
\mathrm{s.t.}\quad &\eqref{digconuser}\nonumber\\
&\|\mathrm{vec}[\sqrt{\mathbf{\eta[k]}}\mathbf{B}[k]^H\mathbf{G}[k]^H\mathbf{W}[k]-\sqrt{\mathbf{\eta[k]}}]\|_F\leq\chi
\end{align}
where $\bf{\eta}[k]=\mathrm{diag}[\eta_1[k],\eta_2[k],...,\eta_U[k]]$ in which $\eta_u[k]=\kappa_u[k]\zeta_u[k]^{\mu_u[k]}$, $\mathbf{B}[k]=\mathrm{diag}[b_1[k],b_2[k],...,b_U[k]]$. As an SOCP problem, \eqref{socp} can be efficiently solved by existing optimization tools \cite{boyd2004convex}. The first and second steps are repeated iteratively and finally achieve a local optimum to \eqref{mmsepro}. In detail, the locally optimal digital precoding algorithm is summarized as displayed in Algorithm 2.

\begin{algorithm}
\caption{Locally Optimal Digital Precoding Algorithm}
\begin{algorithmic}[1]
\REQUIRE $\mathbf{F}$
\STATE $\mathbf{W}[k]^0$ is set randomly and $t=0$
\REPEAT
\STATE With the fixed $\mathbf{W}[k]^t$, update the receiver according to \eqref{receiver} and other necessary factors using \eqref{factors}, respectively.
\STATE Optimize $\mathbf{W}[k]^{t+1}$ as a SOCP problem with the fixed $\{\zeta_u[k]^{t+1},\nu_u[k]^{t+1},b_u[k]^{t+1}\}_{u=1}^U$
\STATE $t\leftarrow t+1$
\UNTIL{$(f(x_{t})-f(x_{t-1}))/f(x_{t-1})<\omega$ for a small $\omega>0$}
\end{algorithmic}
\end{algorithm}

$\mathbf{W}[k]^0$ is set randomly in Algorithm 2. The unnormalized digital beamforming $\mathbf{V}[k]^0$ is first initialized using the uniform random distribution. Note that most random distributions could be used here. Then, to satisfy \eqref{digconuser}, we normalize the initialization of digital precoding as $\mathbf{w}_u[k]^0=\frac{\mathbf{v}_u[k]^0}{\|\mathbf{F}\mathbf{v}_u[k]^0\|_F}$ where $\mathbf{V}[k]^0=[\mathbf{v}_1[k]^0,\mathbf{v}_2[k]^0,...,\mathbf{v}_U[k]^0]$.

Although Algorithm 2 is able to achieve a local optimum for the digital precoding, it is originally designed for narrowband systems. In each iteration of Algorithm 2, it solves an SOCP problem which leads to large complexity. The complexity in each iteration of Algorithm 2 further accumulates to make the total computational period quite long. For a wideband system with a large number of subcarriers, it becomes prohibitive. Therefore, in the rest of this subsection, we would like to propose a suboptimal digital precoding design based on simpler weighted MMSE criterion.

Applying the inequality of weighted arithmetic and geometric means, an upper bound of the objective function in \eqref{mmsepro} is written as
\begin{align}
\prod\limits_{u=1}^U(\xi_u[k])^{z_u}\leq&\sum\limits_{u=1}^Uz_u\xi_u[k]\label{msesum}.
\end{align}
To reduce the computational complexity, we transfer the objective function of problem \eqref{mmsepro} into the sum of MSEs. With the new objective function, the problem becomes:
\begin{align}
\min\limits_{\mathbf{W}[k]}\quad &\sum\limits_{u=1}^Uz_u\xi_u[k],k=1,2,...,K\\
\mathrm{s.t.}\qquad &\eqref{digconuser}.\nonumber
\end{align}

To solve this problem, we propose an iterative algorithm for the digital beamforming based on weighted MMSE. In each iteration of weighted MMSE-based algorithm, we first update $\{b_u[k]\}_{u=1}^U$ according to \eqref{receiver} and then calculate the digital beamforming based on MMSE criterion and normalize the digital beamforming. The corresponding unnormalized MMSE beamforming is given by
\begin{align}
\mathbf{V}[k]=&\left(\mathbf{G}[k]\mathbf{B}[k]\mathbf{Z}^H\mathbf{Z}\mathbf{B}[k]^H\mathbf{G}[k]^H+\frac{\mathrm{Tr}[\mathbf{Z}\mathbf{B}[k]^H\mathbf{B}[k]\mathbf{Z}^H]}{U\vartheta}\mathbf{F}^H\mathbf{F}\right)^{-1}\mathbf{G}[k]\mathbf{B}[k]\mathbf{Z}^H\label{immmse}
\end{align}
where $\mathbf{Z}=\mathrm{diag}[z_1,...,z_U]$. To fulfill the power constraints, the digital precoder is finally normalized as
\begin{align}
\mathbf{w}_u[k]=\frac{\mathbf{v}_u[k]}{\|\mathbf{F}\mathbf{v}_u[k]\|_F}\label{dignor}
\end{align}
where $\mathbf{V}[k]=[\mathbf{v}_1[k],\mathbf{v}_2[k],...,\mathbf{v}_U[k]]$. Therefore, the weighted MMSE-based digital precoding is summarized as in Algorithm 3. It is notable that the method to set $\mathbf{W}[k]^0$ is the same as that in Algorithm 2.

\begin{algorithm}
\caption{Weighted MMSE-based Iterative Digital Precoding Algorithm}
\begin{algorithmic}[1]
\REQUIRE $\mathbf{F}$
\STATE $\mathbf{W}[k]^0$ is set randomly and $t=0$
\REPEAT
\STATE With the fixed $\mathbf{W}[k]^{t}$, update $\{b_u[k]^{t+1}\}_{u=1}^U$ according to \eqref{receiver}
\STATE Calculate $\mathbf{W}[k]^{t+1}$ using \eqref{immmse} and \eqref{dignor} with the fixed $\{b_u[k]^{t+1}\}_{u=1}^U$
\STATE $t\leftarrow t+1$
\UNTIL{$(f(x_{t})-f(x_{t-1}))/f(x_{t-1})<\omega$ for a small $\omega>0$}
\end{algorithmic}
\end{algorithm}

The performance difference between the two digital precoding designs is mainly related to the difference between the weighted arithmetic and geometric means of MSEs. To discover the differences between the weighted arithmetic and geometric means of MSEs, we first define the following factor:
\begin{align}
\iota[k]=\frac{\sum\limits_{u=1}^Uz_u\xi_u[k]-\prod\limits_{u=1}^U(\xi_u[k])^{z_u}}{\prod\limits_{u=1}^U(\xi_u[k])^{z_u}}.\label{4.5.0}
\end{align}
Through mathematical derivations, we have the following theorem on the difference between the weighted arithmetic and geometric means of MSEs.
\begin{myTheo}
The upper bound of $\iota[k]$ is related to the maximal and minimal SINRs at each subcarrier, i.e., $\mathrm{SINR}[k]_{max}$ and $\mathrm{SINR}[k]_{min}$, which is denoted as
\begin{align}
\iota[k]\leq\frac{\left(o[k]-1\right)^2}{8}
\end{align}
where $o[k]=\frac{\mathrm{SINR}[k]_{max}}{\mathrm{SINR}[k]_{min}}$ is the ratio of the maximal and minimal SINRs at subcarrier $k$.
\end{myTheo}
Proof: See Appendix C.

In particular, we additionally present the potential insights in the following remark.

\begin{myRem}
The smaller $o[k]$, the smaller the difference between the weighted arithmetic and geometric means of MSEs. For the case with the same weights for all users, the BS intends to provide similar performance to all users. Then the ratio of maximal and minimal SINRs at the same subcarrier approximately equals to 1 and $\iota[k]$ approximately equals to $0$. Even when $o[k]>1$, $\iota[k]$ is still not very large. As an example, at the same subcarrier, for an SINR difference between the maximal and minimum ones of $10\log_{10}o[k]=2$ dB which is already quite large, the performance difference between the arithmetic and geometric means is very small, with $\iota[k]\approx4.28\%$. Knowing that $\prod\limits_{u=1}^U(\xi_u[k])^{z_u}$ is very small if the transmission is spectrally efficient, the absolute difference between the weighted geometric mean and arithmetic means, i.e., $\sum\limits_{u=1}^Uz_u\xi_u[k]-\prod\limits_{u=1}^U(\xi_u[k])^{z_u}$, should be much smaller. For another example, if the weighted spectral efficiency at subcarrier $k$ is $2$ bits/s/Hz and $10\log_{10}o[k]=2$ dB, the weighted geometric and arithmetic means of MSEs are respectively $\prod\limits_{u=1}^U(\xi_u[k])^{z_u}=0.2500$ and $\sum\limits_{u=1}^Uz_u\xi_u[k]=0.2607$ where only small difference exists. In this case, the weighted MMSE-based iterative digital beamforming has performance similar to that of the locally optimal algorithm.
\end{myRem}

However, it is not rigorous to say that $\iota[k]$ is always small for any case, especially if $o[k]$ is large. It could happen when the ratio of the maximal and minimal weights among users are extremely large. To provide more insights, more discussions will be included in Section V.

\subsection{Alternating Maximization Framework for Hybrid Precoding}

With the designs for the analog and digital beamforming in Algorithms 1-3, we summarize the alternating maximization framework for hybrid precoding as displayed in Algorithm 4. In each loop of Algorithm 4, we first update the analog beamforming using Algorithm 1 with the fixed digital beamforming and then optimize the digital beamforming using Algorithm 2 or 3 with the fixed analog beamforming.

\begin{algorithm}
\caption{Alternating Maximization Framework for Hybrid Precoding}
\begin{algorithmic}[1]
\REQUIRE $\mathbf{F}^{(0)}$, $\mathbf{W}[k]^{(0)}$
\STATE Set $t=0$ and $R_{\rm sum}^{(-1)}=0$
\REPEAT
\STATE With the current $\{\mathbf{W}[k]^t\}_{k=1}^K$, update $\mathbf{F}^{(t+1)}$ using Algorithm 1
\STATE With the current $\mathbf{F}^{(t+1)}$, optimize $\{\mathbf{W}[k]^{(t+1)}\}_{k=1}^K$ using Algorithm 2 or Algorithm 3
\STATE $t\leftarrow t+1$
\UNTIL{$(f(x_{t})-f(x_{t-1}))/f(x_{t-1})<\omega$ for a small $\omega>0$}
\end{algorithmic}
\end{algorithm}

If Algorithm 2 is used for the digital precoding, Algorithm 4 is guaranteed to be locally optimal since Algorithm 1 and 2 are both guaranteed to be locally optimal. For the sake of convenience, we name the locally optimal alternating maximization as locally optimal alternating optimization for hybrid beamforming (LAOHB). In constrast, if Algorithm 3 is employed for the digital beamforming, we name Algorithm 4 as alternating optimization for hybrid beamforming (AOHB). As shown in Theorem 1, the performance of Algorithm 2 and Algorithm 3 is similar. It further implies that the performance of AOHB is also quite similar to that of LAOHB if $o[k]$ is not very large.

\begin{table}
\centering
\caption{Computational Complexities of the Proposed Algorithms.}
\begin{tabular}{|c|c|} %l(left)������ʾ r(right)������ʾ c������ʾ
\hline
Algorithms&Computational complexity per iteration\\
%\hline
%Algorithm 1&$O(mMU)$\\
\hline
Algorithm 2&$\mathcal{O}(KMN_{RF}^2U^{3.5})$\\
\hline
Algorithm 3&$\mathcal{O}(KMN_{RF}^2)$\\
\hline
\end{tabular}
\end{table}

In the alternating maximization framework for hybrid precoding, the analog precoding is optimized using the same algorithm. Thus, the computational complexity in each loop of Algorithm 4 for the analog precoding is the same. The main computational load of Algorithm 2 arises from solving \eqref{receiver} and \eqref{socp}. In \eqref{receiver}, the main computational load can be summarized as three matrix multiplication operations $\mathbf{g}[k]^H=\mathbf{h}[k]^H\mathbf{F}$, $\mathbf{a}_u[k]^H=\mathbf{g}_u[k]^H\mathbf{W}[k]$ and $\mathbf{a}_u[k]^H\mathbf{a}_u[k]$. Their complexities could be respectively evaluated as $\mathcal{O}(MN_{RF})$, $\mathcal{O}(UN_{RF})$ and $\mathcal{O}(U^2)$ \cite{algorithmdesign2008}. Since \eqref{receiver} should be calculated for each user each subcarrier, the complexity over all subcarriers regarding \eqref{receiver} is $\mathcal{O}(KUMN_{RF}+KU^2N_{RF}+KU^3)$. In problem \eqref{socp}, there are $U$ second-order-cone (SOC) constraints where each of them consists of $2M$ real dimensions ($\|\mathbf{F}\mathbf{w}_u[k]\|_F\leq1,u=1,...,U$), one SOC constraint with $2U^2$ real dimensions ($\|\mathrm{vec}[\sqrt{\mathbf{\eta[k]}}\mathbf{B}[k]^H\mathbf{G}[k]^H\mathbf{W}[k]-\sqrt{\mathbf{\eta[k]}}]\|_F\leq\chi$) and $2N_{RF}U+1$ real optimization variables ($\chi$ and $\mathbf{W}\in\mathbb{C}^{N_{RF}\times U}$). From \cite[Section 1.4]{socp1998}, for Algorithm 2, the number of iterations is upper bounded by $\mathcal{O}(\sqrt{U+1})$ and the complexity of each iteration is on the order of $\mathcal{O}((2N_{RF}U+1)^2(2U^2+2MU))$. Therefore, the computational complexity of Algorithm 2 for the whole system is $\mathcal{O}(K\sqrt{U+1}(2N_{RF}U+1)^2(2U^2+2MU)+KUMN_{RF}+KU^2N_{RF}+KU^3)$ in each iteration. Considering $M\gg N_{RF}\geq U$, the complexity of Algorithm 2 per iteration could be concluded as $\mathcal{O}(KMN_{RF}^2U^{3.5})$. For Algorithm 3, the main computational load is solving \eqref{receiver} and \eqref{immmse}. In \eqref{immmse}, the main computational task arises from matrix multiplication and matrix inversion operations. The inversion of a $N_{RF}\times N_{RF}$ matrix has a complexity in the order of $\mathcal{O}(N_{RF}^{2.376})$ \cite{matrixinverse1990}. Using steps similar to those analyzing \eqref{receiver}, we can obtain the complexity of matrix multiplication in \eqref{immmse} is $\mathcal{O}(UN_{RF}^2+MN_{RF}^2+UN_{RF}^2)$. Combining the computational complexities of \eqref{receiver} and \eqref{immmse}, Algorithm 3 requires $\mathcal{O}(KUN_{RF}^2+KMN_{RF}^2+KUN_{RF}^2+KN_{RF}^{2.376}+KUMN_{RF}+KU^2N_{RF}+KU^3)$ operations over all subcarriers at each iteration. Since $M\gg N_{RF}\geq U$, we write the complexity of Algorithm 3 per iteration as $\mathcal{O}(KMN_{RF}^2)$. The computational complexities of Algorithm 2 and 3 are summarized in Table I.

\section{Low-Complexity Closed-form Hybrid Precoding Design}
In this section, we would like to propose one simplified closed-form method with reduced complexity. Although it is not supposed to maximize the system performance, it achieves suboptimal performance employing matrix decomposition for the analog beamforming and weighted MMSE for the digital beamforming. Moreover, this solution can be the initialization for the iterative algorithms. Compared with random initialization, it may help dramatically save time for the iterative algorithms to achieve a locally optimal result.

Since the original problem \eqref{pro} is complicated and non-convex, to propose simple designs, we need to begin with simplifying the problem. To simplify the problem, we would like to avail ourselves of some experience from the designs employed in single carrier systems which approach the performance of the pure digital beamforming using zero-forcing (ZF) precoding in massive MIMO \cite{6928432}. Similarly to the beamforming methods in \cite{7160780, 6928432, 7947159, 8383712}, the hybrid beamforming design for multiuser MIMO could be divided into two stages: (1) First maximize the signal power on the entire band for each user utilizing the analog precoder ignoring the interference among users; (2) Second, design the digital beamformers according to the effective channels using the linear digital beamforming schemes such as MMSE to cope with the multiuser interference. To be brief, the analog beamforming helps reap diversity and the digital beamforming mitigates residual multiuser interference in spatial multiplexing.

In the first stage, we assume that $N_u$ columns of $\mathbf{F}$ serve the $u$-th user to maximize the diversity where $\sum\limits_{k=1}^UN_u=N_{RF}$. The analog beamforming matrix could be represented by
%\begin{align}
$\mathbf{F}=[\mathbf{F}_1,\mathbf{F}_2,...,\mathbf{F}_U]$
%\end{align}
where $\mathbf{F}_u\in\mathbb{C}^{M\times N_u}$. Ignoring the interference and the digital beamforming, the analog precoder design is converted into maximizing the signal power for all users which can be formulated as
\begin{align}
\max\limits_{\{N_u\}_{u=1}^U,\mathbf{F}_u} &\quad\sum\limits_{u=1}^U\sum\limits_{k=1}^K\|\mathbf{h}_u[k]^H\mathbf{F}_u\|_F^2\label{simpro}\\
\mathrm{s.t.}\quad &\eqref{anacon}.\nonumber
\end{align}
Although the problem is simplified now, \eqref{simpro} is still non-convex. To address this issue, we would like to propose a general design, i.e., channel matrix decomposition design (CMDD)-based hybrid precoding. Here, we first solve this problem as conventional beamformer design by dropping the constraint of the analog precoder and then restrict the analog beamformer matrix to satisfy the constant amplitude constraint. Without the constant amplitude constraints, we rewrite problem \eqref{simpro} as
\begin{align}
\max\limits_{\{N_u\}'s,\mathbf{F}_u} &\quad\sum\limits_{u=1}^U\sum\limits_{k=1}^K\|\mathbf{h}_u[k]^H\mathbf{F}_u\|_F^2,\\
\mathrm{s.t.}\qquad &\|\mathbf{f}_n\|_F=1,\forall n.\nonumber
\end{align}
To solve this problem, the steps of the proposed solution are described as follows.

Firstly, we apply eigenvalue decomposition to $\mathbf{H}_u\mathbf{H}^H_u$ as
%\begin{align}
$\mathbf{H}_u\mathbf{H}^H_u=\mathbf{P}_u\mathbf{M}^u\mathbf{P}_u^H$
%\end{align}
where $\mathbf{H}_u=[\mathbf{h}_u[1],\mathbf{h}_u[2],...,\mathbf{h}_u[K]]$ stands for the channel matrix of the $u$-th user at all subcarriers. $\mathbf{M}^u=\mathrm{diag}[\rho_{u1},\rho_{u2},...,\rho_{uM}]$ is a diagonal matrix where $\rho_{u1}\geq\rho_{u2}\geq...\geq\rho_{uM}\geq0$. The $i$-th column of $\mathbf{P}_u$, i.e., $\mathbf{p}_i^u$, is the eigenvector corresponding to $\rho_{ui}$. Then, we easily organize the first to $N_u$-th columns of $\mathbf{P}_u$ as $[\mathbf{P}_u]_{:,1:N_u}$ which leads to
\begin{align}
\sum\limits_{k=1}^K\|\mathbf{h}_u[k]^H[\mathbf{P}_u]_{:,1:N_u}\|_F^2
=&\|\mathbf{H}_u^H[\mathbf{P}_u]_{:,1:N_u}\|_F^2\nonumber\\
=&\mathrm{Tr}[([\mathbf{P}_u]_{:,1:N_u})^H\mathbf{H}_u\mathbf{H}_u^H[\mathbf{P}_u]_{:,1:N_u}]\nonumber\\
=&\sum\limits_{n=1}^{N_u}\rho_{un}^2.
\end{align}
According to \eqref{simpro}, the $\{N_u\}_{u=1}^U$ are decided by
\begin{align}
\max\limits_{\{N_u\}_{u=1}^U}\quad &\sum\limits_{u=1}^{U}\sum\limits_{n=1}^{N_u}\rho_{un}^2\\
\mathrm{s.t.} \quad &N_u\geq 1, \forall u.\label{larger1}
\end{align}
In this way, we obtain an approximated analog beamformer design for each user. However, the solution still does not fulfill the constant amplitude constraints of the analog beamforming. Hence, the remaining step is to solve
\begin{align}
\min\limits_{\mathbf{F}_u}\quad &\|\mathbf{F}_u-[\mathbf{P}_u]_{:,1:N_u}\|_F^2\\
%& \begin{array}{r@{\quad}r@{}l@{\quad}l}
\mathrm{s.t.}\quad &|f_{ni}|=\frac{1}{\sqrt{M}},  i=1,2,\ldots,M.
%\end{array}
\end{align}
By applying the property of the Frobenius norm and trace, we can get
%\begin{align}
%\|\mathbf{f}_u-\mathbf{p}_u^{opt}\|_F^2=2-2\mathrm{Tr}[\Re[\mathbf{f}_u^H\mathbf{p}_u^{opt}]].\label{minF}
%\end{align}
\begin{align}
&\|\mathbf{F}_u-[\mathbf{P}_u]_{:,1:N_u}\|_F^2\nonumber\\
=&\mathrm{Tr}[(\mathbf{F}_u-[\mathbf{P}_u]_{:,1:N_u})^H(\mathbf{F}_u-[\mathbf{P}_u]_{:,1:N_u})]\nonumber\\
%=&\mathrm{Tr}[\mathbf{F}_u^H\mathbf{F}_u+(\hat{\mathbf{P}}_u)^H\hat{\mathbf{P}}_u-\mathbf{F}_u^H\hat{\mathbf{P}}_u-(\hat{\mathbf{P}}_u)^H\mathbf{F}_u]\nonumber\\
=&\mathrm{Tr}[\mathbf{F}_u^H\mathbf{F}_u]+\mathrm{Tr}[([\mathbf{P}_u]_{:,1:N_u})^H[\mathbf{P}_u]_{:,1:N_u}]-\mathrm{Tr}[\mathbf{F}_u^H[\mathbf{P}_u]_{:,1:N_u}]-\mathrm{Tr}[([\mathbf{P}_u]_{:,1:N_u})^H\mathbf{F}_u]\nonumber\\
=&2N_u-2\mathrm{Tr}[\Re[(\mathbf{F}_u)^H[\mathbf{P}_u]_{:,1:N_u}]].\label{minF}
\end{align}
According to \eqref{minF}, the minimum value is achieved when $\mathbf{f}_u$ has the same phase components as $[\mathbf{P}_u]_{:,1:N_u}$, i.e.,
\begin{align}
\mathbf{F}_u(i,j)=\frac{1}{\sqrt{M}}\frac{[\mathbf{P}_u]_{:,1:N_u}(i,j)}{|[\mathbf{P}_u]_{:,1:N_u}(i,j)|}\label{08.14}
\end{align}
where $\mathbf{F}_u(i,j)$ and $[\mathbf{P}_u]_{:,1:N_u}$ are the $(i,j)$-th element of $\mathbf{F}_u$ and $[\mathbf{P}_u]_{:,1:N_u}$, respectively. Thus far, the analog beamforming design is completed.

In the second stage, the digital precoder is designed according to weighted MMSE to manage the multiuser interference. For each subcarrier, the effective channel matrix for all users is defined in \eqref{effectivech}. Then, the base station calculates unnormalized MMSE precoder based on effective channels as
\begin{align}
\mathbf{V}[k]&=\left(\mathbf{G}[k]\mathbf{Z}^H\mathbf{Z}\mathbf{G}[k]^H+\frac{1}{\vartheta}\mathbf{I}\right)^{-1}\mathbf{G}[k]\mathbf{Z}^H.\label{MMSE}
\end{align}
After that, we normalize the digital beamforming according to \eqref{dignor}.

Combining the two stages, we obtain the proposed the low-complexity hybrid beamforming design. The steps of CMDD-based hybrid precoding are summarized as in Algorithm 5.

\begin{algorithm}
\caption{CMDD-based Hybrid Precoding}
\begin{algorithmic}[1]
\STATE {\bf First stage:} Analog beamforming design:
\STATE For each user $u$, $u=1,2,...,U$, BS calculates the eigenvalues of $\mathbf{H}_u\mathbf{H}^H_u$ for user $u$ as $\rho_{u1}\geq...\geq\rho_{uM}$
\STATE Among all users, BS first selects the eigenvectors corresponding to the largest eigenvalue for each user, i.e., $\mathbf{p}_1^u$
\STATE If $N_{RF}>U$, BS selects $N_{RF}-U$ eigenvectors corresponding to the $N_{RF}-U$ biggest eigenvalues among the remaining eigenvectors, i.e., $\{\mathbf{p}_i^u\}_{i,u=2,1}^{M,U}$
\STATE Set all selected eigenvectors into the analog beamforming matrix as $\mathbf{F}=[\mathbf{p}_1^1,...,\mathbf{p}_{N_1}^1,\mathbf{p}_1^2,...,\mathbf{p}_{N_2}^2,...,\mathbf{p}_{N_U}^U]$ and normalize the analog precoding according to \eqref{08.14}
\STATE {\bf Second stage:} Multiuser per-carrier digital beamforming design:
\STATE For each subcarrier $k$, $k=1,2,...,K$, BS calculates weighted MMSE beamforming according to \eqref{MMSE} and \eqref{dignor}
\end{algorithmic}
\end{algorithm}

We also separate the complexity analysis of Algorithm 5 into two stages. In the first stage, the main computational load is the eigenvalue decomposition to $\mathbf{H}_u\mathbf{H}_u^H$ which can be achieved by applying singular value decomposition to $\mathbf{H}_u$. Therefore, the complexity of the first step is $\mathcal{O}(MK^2)$ per user \cite{svd2007}, or $\mathcal{O}(UMK^2)$ for the whole system. Then, for the second stage, the main computational task arises from calculating \eqref{MMSE}. Using steps similar to those analyzing \eqref{receiver}, the second stage requires $\mathcal{O}(2KUN_{RF}^2)$ operations over all subcarriers. Therefore, the overall complexity of the Algorithm 5 is $\mathcal{O}(UMK^2+2KUN_{RF}^2)$. Knowing that $K\gg N_{RF}$ in the OFDM-based system and $M\gg N_{RF}\geq U$, we write the complexity of the Algorithm 5 as $\mathcal{O}(UMK^2)$.

Due to the fact that $K\gg N_{RF}$ in the OFDM-based system and $N_{RF}$ is not much larger than $U$, it probably indicates $UMK^2>KMN_{RF}^2$ which implies that the complexity of Algorithm 5 is more than that of Algorithm 3 per iteration. While comparing the complexities of the Algorithm 5 and that of Algorithm 2 per iteration, it is difficult to compare $KMN_{RF}^2U^{3.5}$ and $K^2MU$ directly. For example, if $M=64$, $K=64$, $N_{RF}=U=4$, we have $KMN_{RF}^2U^{3.5}>K^2MU$ which implies that the single iteration of Algorithm 2 has more complexity than the Algorithm 5; while with the same settings except that $K=1024$, the complexity of Algorithm 2 per iteration is less than that of the Algorithm 5.

\section{Simulation Results}
In this section, we present numerical results for all OFDM-based hybrid beamforming designs proposed in this paper. In this section, it is assumed that all weights are the same, i.e., $z_1=...=z_U=\frac{1}{U}$, unless otherwise stated. Between BS and each user, we take the assumption that there are $5$ clusters and $10$ scatterers exist in each cluster \cite{6834753}. AODs of scatterers in each cluster are generated according to a Laplacian distribution with random mean cluster angles $\bar{\theta}_{uc}\in[0,2\pi)$ and angular spreads of $10$ degrees within each cluster. The method of generating $\tau_{uc}$ is according to \cite{7961152, 8306126}, i.e., $\tau_{uc}$ is uniformly distributed over $[0,DT)$ where $D=8$. $\alpha_{ucl}$ is a random variable following the complex Gaussian distribution with zero mean and unit variance. Moreover, the pulse shaping filter $p(\tau)$ is expressed as
\begin{align}
p(\tau)=\left\{\begin{array}{ll}
1, & -T\leq\tau<0\\
0, & {\rm otherwise}.
\end{array}\right.
\end{align}
LAOHB and the AOHB in Section III are respectively denoted as Hybrid-LAO and Hybrid-AO. The low-complexity hybrid precoding design proposed in Section IV is represented as Hybrid-CMDD. To display the performance differences between our proposed designs and other designs, we also provide the simulation results of the pure digital beamforming using weighted MMSE and the hybrid beamforming design proposed in \cite{7737056}. Using steps similar to those analyzing complexity of Algorithm 5, we can calculate the complexity of method in \cite{7737056} as $\mathcal{O}(MK^2)$. Since $UMK^2>MK^2$, the complexity of the Algorithm 5 is more than that of the method in \cite{7737056}.

%\begin{figure*}
%\begin{minipage}[c]{0.32\textwidth}
%\centering
%\includegraphics[width=0.9\textwidth]{iterations.eps}
%\caption{Convergence characteristics of alternating maximization algorithms for different initializations with $M=64$, $K=64$, $U=N_{RF}=8$ and $\vartheta=10$ dB.}
%\end{minipage}%
%\hfill
%\begin{minipage}[c]{0.32\textwidth}
%\centering
%\includegraphics[width=0.9\textwidth]{R8SNRdB0-20.eps}
%\caption{Average spectral efficiency for different $\{\vartheta\}$'s with $M=64$, $K=64$ and $U=8$.}
%\end{minipage}%
%\hfill
%\begin{minipage}[c]{0.32\textwidth}
%\centering
%\includegraphics[width=0.9\textwidth]{RF8U4-8.eps}
%caption{Sum spectral efficiency per subcarrier for different numbers of users with $M=64$, $K=64$, $N_{RF}=8$ and $\vartheta=10$ dB.}
%\end{minipage}
%\end{figure*}

\begin{figure}[!htpb]
\centering
\begin{minipage}{1\textwidth}
\centering
\includegraphics[width=0.6\textwidth]{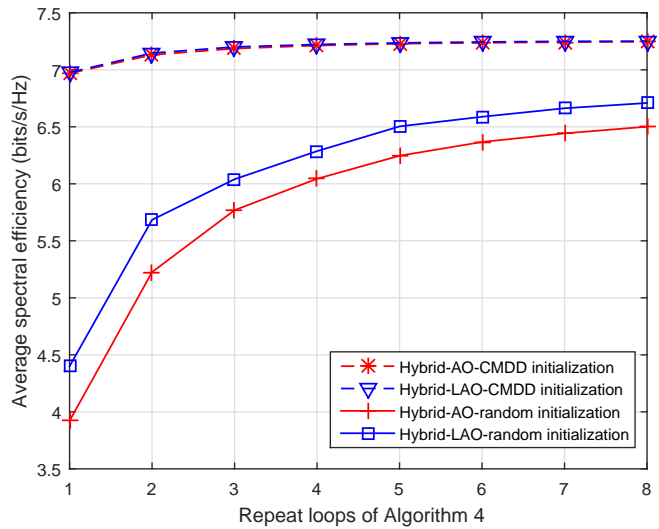}
\caption{Convergence characteristics of alternating maximization algorithms for different initializations with $M=64$, $K=64$, $U=N_{RF}=8$ and $\vartheta=10$ dB.}
\end{minipage}
\end{figure}

In Section III-C, we mathematically analyze the complexities of Algorithms 2 and 3 per iteration. To compare the complexities of the overall algorithms, we also report the running times of LAOHB and AOHB in Table II. ``Times'' in Table II are averaged over $500$ independent channel realizations for the overall alternating maximization framework which does not include the time spent for generation of channel state information. Note that the running time of algorithms are compared under the same conditions of hardware and system settings. In practice, the implementation frequency of this algorithm depends on the coherence time of the physical channels which are in milliseconds. The channel coherence time is commonly defined as the time in which the channel can be considered constant which is not an exact definition. Moreover, the coherence time is generally related to many practical factors, i.e., the multipath, propagation environment, Doppler effect, etc. It should be decided by different practical conditions in different systems. Thus, it is difficult to provide an exact coherence time value in general. On the other hand, the running times provided in Table II are results run by MATLAB in a personal computer. The algorithms run in a practical system would be run by, e.g., FPGA or on chips, whose running times would be strictly restricted and much less than those in Table II. As displayed in Table II, AOHB costs far less running time than LAOHB.

\begin{table}[!htbp]
\centering
\caption{Running time of alternating maximization algorithms with $M=64$, $K=64$, $N_{RF}=U=8$, $\vartheta=10\ {\rm dB}$.}
\begin{tabular}{|c|c|c|}
\hline
Algorithms&LAOHB&AOHB\\
\hline
Average running time&532.5692\ s&2.4625\ s\\
\hline
\end{tabular}
\end{table}

Fig. 1 depicts the convergence characteristics of the alternating maximization algorithms with different initializations. It is shown that the alternating maximization algorithms initialized by the solution to Algorithm 5 (CMDD-based hybrid precoding) converge much faster than when a random initialization is used. With a random initialization, the complexities of LAOHB and AOHB grow since the convergence of Algorithm 4 requires many more repeat loops. Therefore, in the following, the solution to Algorithm 5 will always be employed as the initialization for both alternating maximization algorithms.

\begin{figure}[!htpb]
\centering
\begin{minipage}{1\textwidth}
\centering
\includegraphics[width=0.6\textwidth]{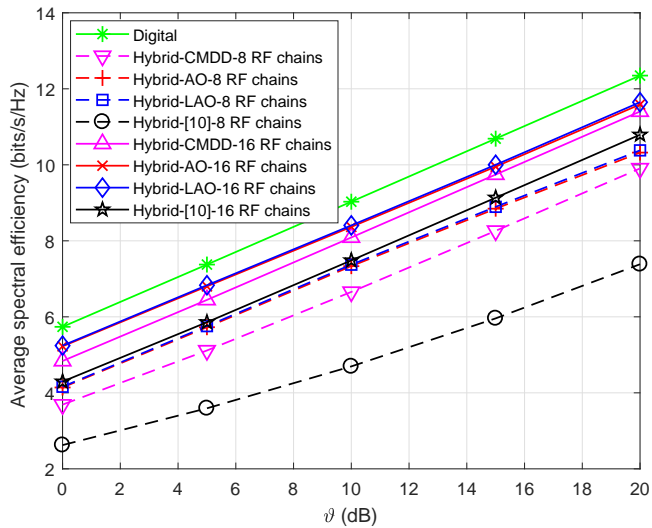}
\caption{Average spectral efficiency for different $\{\vartheta\}$'s with $M=64$, $K=64$ and $U=8$.}
\end{minipage}
\end{figure}

We investigate the performance of all proposed designs versus $\vartheta$ in Fig. 2. As observed from the figure, the pure digital beamforming which serves as the benchmark for comparison with hybrid beamforming designs achieves the highest performance. The two alternating optimization algorithms have similar performance and exhibit the smallest gap compared to the pure digital beamforming. However, the performance of the low-complexity method proposed in this paper is only $1$ or $2$ dB away from that of the alternating optimization algorithms depending on whether we consider $16$ or $8$ RF chains. Moreover, the gains of alternating maximization algorithms are not large when $N_{RF}=16>K$. When the number of RF chains is limited, the dimension of digital beamforming matrix is also limited. In this case, the function of digital beamforming is also limited and system performance is dominated by the analog beamforming. As $N_{RF}$ increases, the impact of digital beamforming becomes more significant while the impact of analog beamforming fades. Since LAOHB and AOHB make much effort on optimizing analog beamforming, they are more suitable for $N_{RF}$ limited scenario but they do not show large gains over CMDD initialization when the number of RF chains is large.

\begin{figure}[!htpb]
\centering
\begin{minipage}{1\textwidth}
\centering
\includegraphics[width=0.6\textwidth]{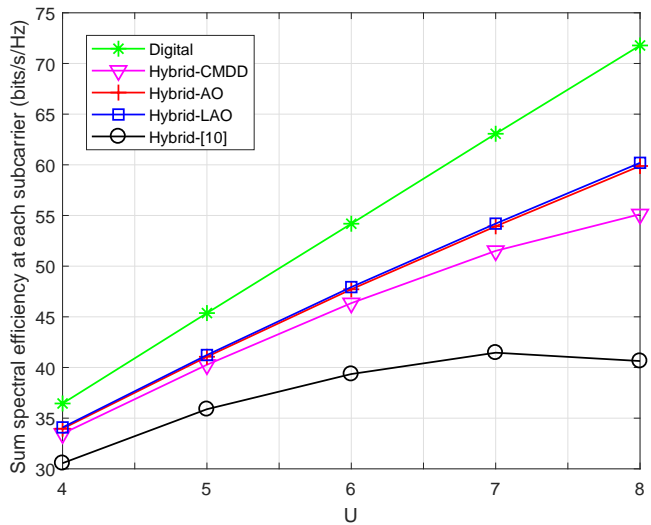}
\caption{Sum spectral efficiency per subcarrier for different numbers of users with $M=64$, $K=64$, $N_{RF}=8$ and $\vartheta=10$ dB.}
\end{minipage}
\end{figure}

In Fig. 3, we compare the average spectral efficiencies versus the number of users. The pure digital beamforming outperforms the alternating optimization algorithms and the alternating optimization algorithms outperform the low-complexity hybrid beamforming design. As displayed in the figure, all designs have similar performance when $U$ is small. With the increase of $U$, the performance gap between the pure digital beamforming and all hybrid beamforming designs becomes larger. The larger $\frac{U}{N_{RF}}$ is, the larger the gaps between the pure digital beamforming and hybrid beamforming designs become, and the larger the gaps between the low-complexity hybrid beamforming and the alternating optimization algorithms are. The dimension of the analog beamforming does not grow with $U$, but that of digital beamforming does. %Therefore, the impact of hybrid beamforming with analog beamforming can not maintain a satisfying level similarly to the pure digital beamforming.
Therefore, the spectral efficiency growth of hybrid beamforming with analog beamforming can not be as large as that of the pure digital beamforming.

\begin{figure}[!htpb]
\centering
\begin{minipage}{1\textwidth}
\centering
\includegraphics[width=0.6\textwidth]{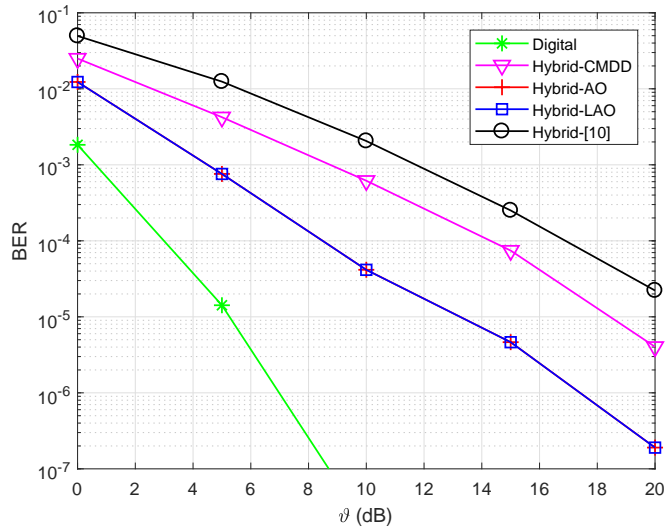}
\caption{Average BER for different $\{\vartheta\}$'s with $M=64$, $K=64$, $U=8$ and $N_{RF}=12$.}
\end{minipage}
\end{figure}

In Fig. 4, we provide the bit error rate (BER) performance using $16$-quadrature amplitude modulation (QAM) for different beamforming designs. The results in the figure also confirm that the two proposed alternating maximization algorithms have the smallest performance gaps compared to the pure digital beamforming. As $\vartheta$ increases, the interference management which is related to channels at different subcarriers becomes more and more significant than improving signal power to the system performance. Over frequency selective channels, frequency selective digital beamforming tends to perform better than the hybrid beamforming using the frequency flat analog beamforming, especially for large $\vartheta$.

\begin{figure}[!htpb]
\centering
\begin{minipage}{1\textwidth}
\centering
\includegraphics[width=0.6\textwidth]{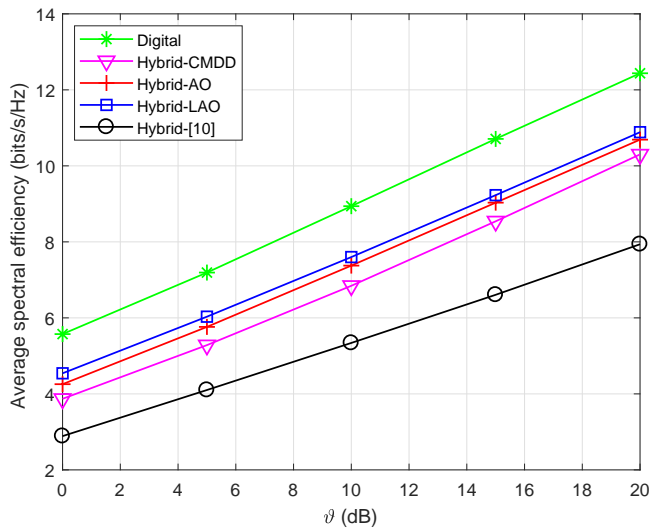}
\caption{Average spectral efficiency for different $\{\vartheta\}$'s with $M=64$, $K=64$, $U=N_{RF}=4$.}
\end{minipage}
\end{figure}

We also provide some results in Fig. 5 with different weights for different users with $z_1=0.7$ and $z_2=z_3=z_4=0.1$. Different from the case with the same weights for all users, the gap between LAOHB and AOHB becomes larger and not negligible. It confirms the idea under Theorem 1 that the $\{\iota[k]\}'s$ could become large when the ratio of the maximal and minimal weights among users are extremely large which further causes the performance difference between LAOHB and AOHB to become larger.

In massive MIMO, the analog precoding is based on the high dimensional physical channel which challenges the availability of perfect channel information. However, the digital precoding is based on the low dimensional effective channel which makes channel estimation easier. Here, we also would like to provide some simulation results on the performance of the hybrid precoding which is designed based on the channel information with random error. By introducing a channel error, we characterize the estimated instantaneous channel state information as \cite{6375940}
\begin{align}
\hat{\mathbf{h}}_u[k]=\varsigma_u^h[k]\mathbf{h}_u[k]+\sqrt{1-\varsigma_u^h[k]^2}\mathbf{e}_u^h[k]
\end{align}
where $\varsigma_u^h[k]\in[0,1]$ refers to the channel state information accuracy level and $\mathbf{e}_u^h[k]\sim\mathcal{CN}(\mathbf{0}_M,\mathbf{I})$ is the error vector. Similarly, the estimated effective channel state information is written as
\begin{align}
\hat{\mathbf{g}}_u[k]=\varsigma_u^g[k]\mathbf{g}_u[k]+\sqrt{1-\varsigma_u^g[k]^2}\frac{\|\mathbf{g}_u[k]\|_F}{N_{RF}}\mathbf{e}_u^g[k]
\end{align}
where $\varsigma_u^g[k]\in[0,1]$ denotes the effective channel state information accuracy level and $\mathbf{e}_u^g[k]\sim\mathcal{CN}(\mathbf{0}_M,\mathbf{I})$ stands for the error vector.

\begin{figure}[!htpb]
\centering
\begin{minipage}{1\textwidth}
\centering
\includegraphics[width=0.6\textwidth]{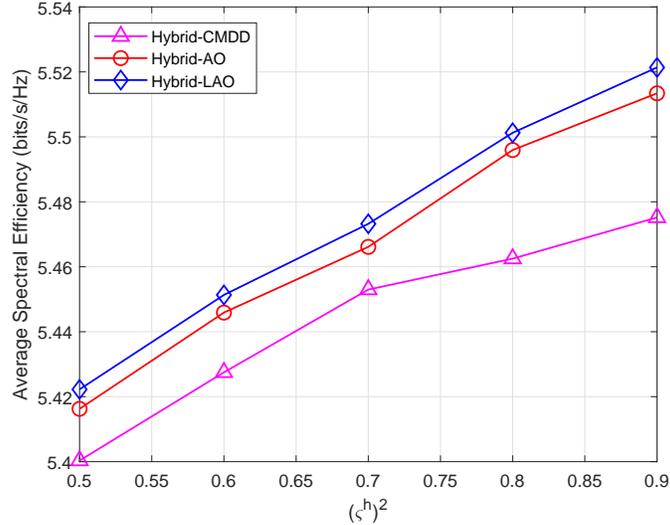}
\caption{Average spectral efficiency for different $\{(\varsigma^h)^2\}$'s with $M=64$, $K=64$, $N_{RF}=8$, $U=4$, $(\varsigma^g)^2=0.95$ and $\vartheta=20$ dB.}
\end{minipage}
\end{figure}

Fig. 6 compares three designs proposed in this paper with different $\{\varsigma^h\}$'s and $\{\varsigma^g\}$'s by taking the assumption that $\varsigma_1^h[1]=...=\varsigma_u^h[k]...=\varsigma_U^h[K]=\varsigma^h$ and $\varsigma_1^g[1]=...=\varsigma_u^g[k]...=\varsigma_U^g[K]=\varsigma^g$. It is obvious that varying $\varsigma^h$ only causes minor performance deterioration.

In practical applications, it is possible to adapt Algorithm 5 to different designs for allocating the number of eigenvectors according to the weighting factors. Strategies to choose the number of eigenvectors could be designed specifically for various scenarios. We provide some simulation results to compare performance using different strategies with $z_1=0.4$ and $z_2=z_3=z_4=0.2$ as shown in Fig. 7. In Strategy 1, the BS allocates RF chains for different users as $N_1=N_2=N_3=N_4=2$; while for Strategy 2, we utilize $N_1=5$ and $N_2=N_3=N_4=1$. It is shown that trade-off between performance and fairness could be achieved by using different allocating strategies.

\begin{figure}[!htpb]
\centering
\begin{minipage}{1\textwidth}
\centering
\includegraphics[width=0.6\textwidth]{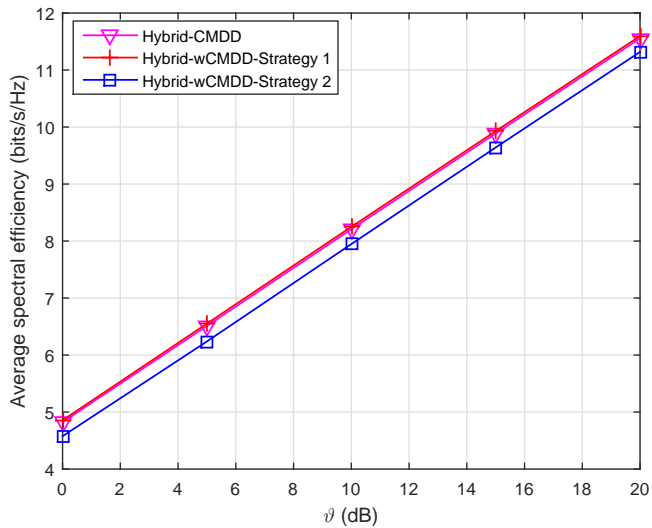}
\caption{Average spectral efficiency for different $\{\vartheta\}$'s with $M=64$, $K=64$, $N_{RF}=8$, $U=4$.}
\end{minipage}
\end{figure}

\section{Conclusion}
In this paper, we studied the optimization of wideband hybrid beamforming for OFDM-based massive MIMO systems, aimed at maximizing the weighted spectral efficiency. Combining a Riemannian manifold optimization algorithm and a locally optimal digital precoding algorithm, we propose a locally optimal alternating maximization algorithm, i.e., LAOHB. To reduce the complexity, we also propose another alternating maximization algorithm, i.e., AOHB, using the Riemannian manifold optimization algorithm for the analog beamforming and a weighted MMSE-based algorithm for the digital beamforming. It is proved that LAOHB and AOHB have similar performance if the ratio of the maximal and minimal weights is not very large. Moreover, a low-complexity closed-form hybrid beamforming design is also proposed. Although the non-iterative design is not supposed to maximize the spectral efficiency, it achieves a suboptimal performance without iteration at low complexity and appears to be a good initialization for iterative algorithms thereby saving iterations.

\appendices
\section{Proof of \eqref{anader}}

To derive $\frac{\partial f(\mathbf{x})}{\partial\mathbf{F}^*}=\sum\limits_{k=1}^K\sum\limits_{u=1}^U\frac{\partial R_u[k]}{\partial\mathbf{F}^*}$, we need to first investigate $\frac{\partial R_u[k]}{\partial\mathbf{F}^*}$ since $f(\mathbf{x})=\sum\limits_{k=1}^K\sum\limits_{u=1}^UR_u[k]$. From \eqref{se}, it is not difficult to achieve
\begin{align}
\frac{\partial R_u[k]}{\partial \mathbf{F}^*}=&\frac{\partial\log_2\left(1+\frac{\vartheta|\mathbf{h}_u[k]^H\mathbf{F}\mathbf{w}_u[k]|^2}{\vartheta\sum\limits_{i\neq u}|\mathbf{h}_u[k]^H\mathbf{F}\mathbf{w}_i[k]|^2+1}\right)}{\partial \mathbf{F}^*}\nonumber\\
=&\frac{\partial\log_2\left(\frac{1+\vartheta\sum\limits_{i=1}^U|\mathbf{h}_u[k]^H\mathbf{F}\mathbf{w}_i[k]|^2}{1+\vartheta\sum\limits_{i\neq u}|\mathbf{h}_u[k]^H\mathbf{F}\mathbf{w}_i[k]|^2}\right)}{\partial \mathbf{F}^*}\nonumber\\
%=&\frac{\partial\log_2\left(1+\vartheta\sum\limits_{i=1}^U|\mathbf{h}_u[k]^H\mathbf{F}\mathbf{w}_i[k]|^2\right)}{\partial \mathbf{F}^*}\nonumber\\
%&-\frac{\partial\log_2\left(1+\vartheta\sum\limits_{i\neq u}|\mathbf{h}_u[k]^H\mathbf{F}\mathbf{w}_i[k]|^2\right)}{\partial \mathbf{F}^*}\nonumber\\
=&\frac{\partial\log_2\left(1+\vartheta\|\mathbf{h}_u[k]^H\mathbf{F}\mathbf{W}[k]\|_F^2\right)}{\partial \mathbf{F}^*}
-\frac{\partial\log_2\left(1+\vartheta\|\mathbf{h}_u[k]^H\mathbf{F}\mathbf{W}_{\bar{u}}[k]\|_F^2\right)}{\partial \mathbf{F}^*}.\label{ap1}
\end{align}
Calculating $\frac{\partial R_u[k]}{\partial\mathbf{F}^*}$ could be separated into deriving the gradient of the first and second terms in \eqref{ap1}. We first focus on the part regarding the first term.

Applying the basic knowledge on differentiation, it is easy to write
\begin{align}
&\frac{\partial \log_2\left(1+\vartheta\|\mathbf{h}_u[k]^H\mathbf{F}\mathbf{W}[k]\|_F^2\right)}{\partial \mathbf{F}^*}
=\frac{1}{\ln2}\frac{\vartheta}{1+\vartheta\|\mathbf{h}_u[k]^H\mathbf{F}\mathbf{W}[k]\|_F^2}\frac{\partial \|\mathbf{h}_u[k]^H\mathbf{F}\mathbf{W}[k]\|_F^2}{\partial \mathbf{F}^*}.\label{ap2}
\end{align}
Then, the remaining steps are to derive $\frac{\partial \|\mathbf{h}_u[k]^H\mathbf{F}\mathbf{W}[k]\|_F^2}{\partial \mathbf{F}^*}$. Using the relationship between trace and Frobenius norm, it yields
\begin{align}
\frac{\partial \|\mathbf{h}_u[k]^H\mathbf{F}\mathbf{W}[k]\|_F^2}{\partial \mathbf{F}^*}
=&\frac{\partial \mathrm{Tr}[\mathbf{h}_u[k]^H\mathbf{F}\mathbf{W}[k]\mathbf{W}[k]^H\mathbf{F}^H\mathbf{h}_u]}{\partial \mathbf{F}^*}\nonumber\\
\overset{(a)}{=}&\frac{\partial \mathrm{Tr}[\mathbf{F}\mathbf{W}[k]\mathbf{W}[k]^H\mathbf{F}^H\mathbf{h}_u[k]\mathbf{h}_u[k]^H]}{\partial \mathbf{F}^*}\nonumber\\
\overset{(b)}{=}&\mathbf{h}_u[k]\mathbf{h}_u[k]^H\mathbf{F}\mathbf{W}[k]\mathbf{W}[k]^H\label{ap3}
\end{align}
where (a) is due to the fact that the trace is invariant under cyclic permutations and (b) utilizes the results in \cite[Table IV]{4203075}. Substituting \eqref{ap3} and \eqref{ap2}, we acquire
\begin{align}
&\frac{\partial \log_2\left(1+\vartheta\|\mathbf{h}_u[k]^H\mathbf{F}\mathbf{W}[k]\|_F^2\right)}{\partial \mathbf{F}^*}
=\frac{1}{\ln2}\frac{\vartheta\mathbf{h}_u[k]\mathbf{h}_u[k]^H\mathbf{F}\mathbf{W}[k]\mathbf{W}[k]^H}{1+\vartheta\|\mathbf{h}_u[k]^H\mathbf{F}\mathbf{W}[k]\|_F^2}.\label{ap4}
\end{align}

Analogously, we further have
\begin{align}
&\frac{\partial \log_2\left(1+\vartheta\|\mathbf{h}_u[k]^H\mathbf{F}\mathbf{W}_{\bar{u}}[k]\|_F^2\right)}{\partial \mathbf{F}^*}
=\frac{1}{\ln2}\frac{\vartheta\mathbf{h}_u[k]\mathbf{h}_u[k]^H\mathbf{F}\mathbf{W}_{\bar{u}}[k]\mathbf{W}_{\bar{u}}[k]^H}{1+\vartheta\|\mathbf{h}_u[k]^H\mathbf{F}\mathbf{W}_{\bar{u}}[k]\|_F^2}.\label{ap5}
\end{align}

Combining \eqref{ap1}, \eqref{ap4}, \eqref{ap5} and $\frac{\partial f(\mathbf{x})}{\partial\mathbf{F}^*}=\sum\limits_{k=1}^K\sum\limits_{u=1}^U\frac{\partial R_u[k]}{\partial\mathbf{F}^*}$, the proof of \eqref{anader} completes.

\section{Derivations on \eqref{factors}}

Due to the fact that \eqref{mmsepro} is nonconvex, we first introduce the new variables $\{\nu_u[k]\}_{u=1}^U$ and transfer the problem into \cite[Lemma 1]{6104172}
\begin{align}
\min\limits_{\{\nu_u[k]\}_{u=1}^U,\mathbf{W}[k]}\quad &\left(\frac{1}{U}\sum\limits_{u=1}^U\nu_u[k]\xi_u[k]^{z_u}\right)^U\label{R.3.1.1}\\
\mathrm{s.t.}\qquad &\eqref{digconuser}\nonumber\\
&\prod\limits_{u=1}^U\nu_u[k]=1,\nu_u[k]>0,\forall u\label{R.3.1.2}.
\end{align}
The optimal $\{\nu_u[k]\}_{u=1}^U$ of this problem is given by \cite[Appendix C]{6104172}
\begin{align}
\nu_u[k]=&\frac{\left[\prod\limits_{i=1}^U\xi_i[k]^{z_u}\right]^{\frac{1}{U}}}{\xi_u[k]^{z_u}}\label{R.3.1.3}.
\end{align}
Then, for any function $f(x)>0$, we have the fact that $\min_x(qf(x))^r$ is equivalent to $\min_xf(x)$ for any $q>0$, $f(x)>0$, $\forall x$ and integer $r>0$ \cite{boyd2004convex}. Since it is not difficult to find $\left(\frac{1}{U}\sum\limits_{u=1}^U\nu_u[k]\xi_u[k]^{z_u}\right)^U>0$ and $\nu_u[k]>0$, we can replace the objective function of \eqref{R.3.1.1} by $\sum\limits_{u=1}^U\nu_u[k]\xi_u[k]^{z_u}$ which implies that \eqref{R.3.1.1} becomes
\begin{align}
\min\limits_{\{\nu_u[k]\}_{u=1}^U,\mathbf{W}[k]}\quad &\sum\limits_{u=1}^U\nu_u[k]\xi_u[k]^{z_u}\label{R.3.1.4}\\
\mathrm{s.t.}\qquad &\eqref{digconuser},\eqref{R.3.1.2}.
\end{align}
It is notable that getting the suboptimal solution of \eqref{R.3.1.4} is not trivial due to terms $\{\nu_u[k]\xi_u[k]^{z_u}\}_{u=1}^U$ in \eqref{R.3.1.4}. To further simplify the problem, we would like to use the fact that $\min\limits_{\zeta>0}\kappa\left(\frac{\nu^\gamma}{\zeta}+\xi\zeta^\mu\right)=\nu\xi^z$ holds true for any strictly positive numbers $\nu$ and $\xi$, and $0<z<1$ where $\gamma=\frac{1}{1-z}$, $\mu=\frac{1}{z}-1$, $\kappa=z\mu^{1-z}$ \cite[Lemma 2]{6104172}. It implies that $\{\nu_u[k]\}_{u=1}^U, \mathbf{W}[k]$ of \eqref{R.3.1.4} can be optimized by solving
\begin{align}
\min\limits_{\substack{\{\zeta_u[k],\nu_u[k]\}_{u=1}^U\\\mathbf{W}[k]}}&\sum\limits_{u=1}^U\kappa_u\left[\frac{\nu_u[k]^{\gamma_u}}{\zeta_u[k]}+\xi_u[k]\zeta_u[k]^{\mu_u}\right]\label{R.3.1.6}\\
s.t.\qquad &\eqref{digconuser},\eqref{R.3.1.2}\\
&\zeta_u[k]>0,\forall u.\label{R.3.1.7}
\end{align}
where
\begin{align}
\gamma_u=&\frac{1}{1-z_u},\mu_u=\frac{1}{z_u}-1,\kappa_u=z_u\mu_u^{1-z_u}\label{R.3.1.8}.
\end{align}
With fixed $\{\nu_u[k]\}_{u=1}^U$ and $\mathbf{W}[k]$, the optimal $\{\zeta_u[k]\}_{u=1}^U$ of \eqref{R.3.1.6} can be obtained by applying the first order partial derivative of the objective function in \eqref{R.3.1.6} and are given as
\begin{align}
\zeta_u[k]=\left[\frac{\nu_u[k]^{\gamma_u}}{\mu_u\xi_u[k]}\right]^{\frac{1}{\mu_u+1}}.\label{R.3.1.9}
\end{align}
In the first step of each iteration in Algorithm 2, we should update factors according to \eqref{R.3.1.3}, \eqref{R.3.1.8} and \eqref{R.3.1.9} which are summaried by \eqref{factors}.

\section{Proof of Theorem 1}
For convenience's sake, we first define the MSEs at subcarrier $k$ for all users in the descending order as $r_1[k]\geq...\geq r_U[k]\geq 0$ and $z_u[k]'$ is the weighting factor corresponding to $r_u[k]$. By applying \cite[Theorem]{cartwright1978refinement}, we get
\begin{align}
\sum\limits_{u=1}^Uz_u\xi_u[k]-\prod\limits_{u=1}^U(\xi_u[k])^{z_u}\leq\frac{1}{2r_{min}[k]}\mathbb{V}ar_z(\Xi)\label{4.5.1}
\end{align}
where $r_{min}[k]$ stands for the minimum value among  $\{r_u[k]\}_{u=1}^U$ and
\begin{align}
\mathbb{V}ar_z(\Xi[k])=&\sum\limits_{u=1}^Uz_u\left(\xi_u[k]-\sum\limits_{u=1}^Uz_u\xi_u[k]\right)^2\nonumber\\
=&\sum\limits_{u=1}^Uz_u[k]'\left(r_u[k]-\bar{r}[k]\right)^2\label{4.5.2}
\end{align}
in which $\bar{r}[k]=\sum\limits_{u=1}^Uz_u[k]'r_u[k]$.

To discover the upper bound of $\mathbb{V}ar_z(\Xi[k])$, we first investigate the problem maximizing the $\mathbb{V}ar_z(\Xi[k])$ as
\begin{align}
\max\limits_{\{r_u[k]\}_{u=1}^U}\quad &\sum\limits_{u=1}^Uz_u[k]'\left(r_u[k]-\bar{r}[k]\right)^2\\
\mathrm{s.t.}\qquad &r[k]_{min}\leq r_u[k]\leq r[k]_{max}, \forall u
\end{align}
where $r[k]_{max}$ refers to the maximum value among  $\{r_u[k]\}_{u=1}^U$. Motivated by the Karush-Kuhn-Tucker (KKT) conditions, we define the Lagrangian:
\begin{align}
%L(\{r_u[k],\epsilon_u[k],\psi_u[k]\}_{u=1}^U)
L=&\sum\limits_{u=1}^Uz_u[k]'\left(r_u[k]-\bar{r}[k]\right)^2+\sum\limits_{u=1}^U\epsilon_u[k]r_u[k]-\sum\limits_{u=1}^U\psi_u[k](r_u[k]-1).
\end{align}
By letting %$\frac{\partial L(\{r_u[k],\epsilon_u[k],\psi_u[k]\}_{u=1}^U)}{\partial r_u[k]}=0$
$\frac{\partial L}{\partial r_u[k]}=0$, we get the necessary conditions, i.e., $r_u[k]=r[k]_{min}$ or $r_u[k]=r[k]_{max}$ or $r_u[k]=\bar{r}[k]$. According to the definition of the weighted average, we %easily
know
\begin{align}
&\sum\limits_{u=1}^{u_1}z_u[k]'r[k]_{max}+\sum\limits_{u=u_1+1}^{U-u_2}z_u[k]'\bar{r}+\sum\limits_{u=U-u_2+1}^{U}z_u[k]'r[k]_{min}=\bar{r}
\end{align}
where $u_1$ is the number of MSEs equivalent to $r[k]_{max}$ and $u_2$ is the number of MSEs equivalent to $r[k]_{min}$. It further implies that
\begin{align}
\bar{r}[k]=&\frac{\sum\limits_{u=1}^{u_1}z_u[k]'r[k]_{max}+\sum\limits_{u=U-u_2+1}^{U}z_u[k]'r[k]_{min}}{\sum\limits_{u=1}^{u_1}z_u[k]'+\sum\limits_{u=U-u_2+1}^{U}z_u[k]'}\nonumber\\
=&\frac{A_1r[k]_{max}+A_2r[k]_{min}}{A_1+A_2}\label{4.5.5}
\end{align}
where $A_1=\sum\limits_{u=1}^{u_1}z_u[k]'$ and $A_2=\sum\limits_{u=U-u_2+1}^{U}z_u[k]'$. Substituting \eqref{4.5.5} into \eqref{4.5.2}, it yields
\begin{align}
\mathbb{V}ar_z(\Xi[k])=&A_1(r[k]_{max}-\bar{r}[k])^2+A_2(r[k]_{min}-\bar{r}[k])^2\nonumber\\
%=&\frac{A_1A_2^2(r[k]_{max}-r[k]_{min})^2+A_1^2A_2(r[k]_{max}-r[k]_{min})^2}{(A_1+A_2)^2}\nonumber\\
=&\frac{A_1A_2(r[k]_{max}-r[k]_{min})^2}{A_1+A_2}\nonumber\\
\overset{(a)}{\leq}&\frac{A_1+A_2}{4}(r[k]_{max}-r[k]_{min})^2\nonumber\\
\leq&\frac{(r[k]_{max}-r[k]_{min})^2}{4}\label{4.5.7}
\end{align}
where (a) is acquired by using the inequality of arithmetic and geometric means. Combining \eqref{4.5.0}, \eqref{4.5.1} and \eqref{4.5.7}, we obtain
\begin{align}
\iota[k]\leq&\frac{(r[k]_{max}-r[k]_{min})^2}{8r[k]_{min}\prod\limits_{u=1}^U(\xi_u[k])^{z_u}}\nonumber\\
\leq&\frac{(r[k]_{max}-r[k]_{min})^2}{8r[k]_{min}^2}\nonumber\\
=&\frac{\left(\frac{1}{r[k]_{min}}-\frac{1}{r[k]_{max}}\right)^2}{8\frac{1}{r[k]_{max}^2}}\nonumber\\
\leq&\frac{\left(\left(\frac{1}{r[k]_{min}}-1\right)-\left(\frac{1}{r[k]_{max}}-1\right)\right)^2}{8\left(\frac{1}{r[k]_{max}}-1\right)^2}\nonumber\\
=&\frac{\left(\mathrm{SINR}[k]_{max}-\mathrm{SINR}[k]_{min}\right)^2}{8\mathrm{SINR}[k]_{min}^2}\nonumber\\
=&\frac{\left(o[k]-1\right)^2}{8}
\end{align}
where $\mathrm{SINR}[k]_{max}$ and $\mathrm{SINR}[k]_{min}$ are respectively used  to denote the SINR corresponding to $r[k]_{min}$ and $r[k]_{max}$.
%\section*{Acknowledgement}

%The authors would like to thank F.R.S.-FNRS for funding the EOS program (EOS project 30452698), INNOVIRIS for funding the COPINE-IOT project and UCL for funding the ARC SWIPT project.

\ifCLASSOPTIONcaptionsoff
  \newpage
\fi

\bibliographystyle{ieeetr}
\bibliography{myreference}

\end{document}